\documentclass[useAMS,usenatbib]{mn2e}
\def\kms{km\,s$^{-1}$}

\def\Ha{H$\alpha$}
\def\Hb{H$\beta$}
\def\Hg{H$\gamma$}

\def\M{M$_{\odot}$}

\usepackage{graphicx}

\title[CSS121015]
{The supernova CSS121015:004244+132827: a clue for understanding super-luminous supernovae.}

\author[Benetti et al.]{S. Benetti$^{1}$\thanks{E-mail: stefano.benetti@oapd.inaf.it}
M. Nicholl$^{2}$, E. Cappellaro$^1$, A. Pastorello$^1$, S. J. Smartt$^2$, N. Elias-Rosa$^{3,1}$,
\newauthor{A. J. Drake$^4$, L. Tomasella$^1$, M. Turatto$^1$, A.~Harutyunyan$^5$, S. Taubenberger$^6$,}
\newauthor{S. Hachinger$^{7,1}$, A. Morales-Garoffolo$^3$, T.-W. Chen$^2$, S.G. Djorgovski$^4$, M. Fraser$^2$,}
\newauthor{A. Gal-Yam$^8$, C. Inserra$^2$, P. Mazzali$^{9,1}$, M. L. Pumo$^1$, J. Sollerman$^{10,11}$, S. Valenti$^{12,13}$,}
\newauthor{D. R. Young$^2$, M. Dennefeld$^{14}$, L. Le Guillou$^{15}$, M. Fleury$^{15}$, P.-F. L\'eget$^{16}$}\\
$^1$INAF - Osservatorio Astronomico di Padova, vicolo dell'Osservatorio 5, I-35122 Padova, Italy \\ 
$^2$Queen's University Belfast, Belfast BT7 1NN, United Kingdom\\
$^3$Institut de Ci\`encies de l'Espai (CSIC-IEEC), Facultat de Ci\`encies, Campus UAB, 08193 Bellaterra, Spain\\
$^4$California Institute of Technology, 1200 E. California Blvd., CA 91225, USA\\
$^5$Telescopio Nazionale Galileo, Fundaci\'on Galileo Galilei - INAF, Rambla Jos\'e Ana Fern\'andez P\'erez, 7, 38712 Bre\~na Baja, TF - Spain \\
$^6$Max-Planck-Institut f\"ur Astrophysik, Karl Schwarzschild Str. 1, 85741 Garching bei M\"unchen, Germany\\
$^7$Institut f{\"u}r Theoretische Physik und Astrophysik, Universit{\"at} W{\"u}rzburg, D-97074 W{\"u}rzburg, Germany\\
$^9$Liverpool John Moores, Liverpool Science Park, IC2 building, 146 Brownlow Hill, Liverpool L3 5RF, UK\\
$^8$Benoziyo Center for Astrophysics, Weizmann Institute of Science, Rehovot 76100, Israel\\
$^{10}$Department of Astronomy, AlbaNova Science Center, Stockholm University, SE-106 91 Stockholm, Sweden\\
$^{11}$The Oskar Klein Centre, AlbaNova, SE-106 91 Stockholm, Sweden\\
$^{12}$Department of Physics, University of California, Santa Barbara, Broida Hall, Mail Code 9530, Santa Barbara, CA 93106-9530, USA\\
$^{13}$Las Cumbres Observatory, Global Telescope Network, 6740 Cortona Drive Suite 102, Goleta, CA 93117, USA\\
$^{14}$CNRS, Institut dÕAstrophysique de Paris (IAP) and University P. et M. Curie (Paris 6), 98bis, Boulevard Arago F-75014 Paris, France \\
$^{15}$UPMC Univ. Paris 06, UMR 7585, Laboratoire de Physique Nucleaire et des Hautes Energies (LPNHE), F-75005 Paris, France\\
$^{16}$Clermont Universit\'e, Universit\'e Blaise Pascal, CNRS/IN2P3, Laboratoire de Physique Corpusculaire, BP 10448, F-63000 \\
~~CLERMONT-FERRAND, France\\
}

\date{Received ................; accepted ................}

\begin{document}

\maketitle

\begin{abstract}
We present optical photometry and spectra of the super luminous type II/IIn supernova CSS121015:004244+132827 (z=0.2868) spanning epochs from -30 days (rest frame) to more than 200 days after maximum. CSS121015 is one of the more luminous supernova ever found and one of the best observed. The photometric evolution is characterized by a relatively fast rise to maximum ($\sim40$ days in the SN rest frame), and by a linear post-maximum decline. The light curve shows no sign of a break to an exponential tail.\\
A broad \Ha~ is first detected at $\sim +40$ days (rest-frame). Narrow, barely-resolved Balmer and [O III] 5007 \AA~lines, with decreasing strength, are visible along the entire spectral evolution.\\
The spectra are very similar to other super luminous supernovae (SLSNe) with hydrogen in their spectrum, and also to SN~2005gj, sometimes considered a type Ia interacting with H-rich CSM. The spectra are also similar to a subsample of H-deficient SLSNe.\\
We propose that the properties of CSS121015 are consistent with the interaction of the ejecta with a massive, extended, opaque shell, lost by the progenitor decades before the final explosion, although a magnetar powered model cannot be excluded. Based on the similarity of CSS121015 with other SLSNe (with and without H), we suggest that the shocked-shell scenario should be seriously considered as a plausible model for both types of SLSN.

\end{abstract} 
 

\section{Introduction} \label{int}

Supernovae (SNe), the dramatic and violent end-points of stellar evolution, lie at the heart of some of the most important problems of modern astrophysics. New observational evidence, largely from extensive transient searches \citep[e.g. the Catalina Real-time Transient Survey, the Palomar Transient Factory, PanSTARRS1, La Silla Quest, etc..][respectively]{dra09,law09,ton12,bal13}, is shedding new light both at the bright and faint ends of the SN luminosity distribution. In particular, a number of exceptionally luminous objects (SLSNe) are challenging our existing theoretical picture of the late evolution and explosion of massive stars.
It is crucial to understand the nature of these events because, with the next generation of instruments, they may be used to probe star formation and chemical enrichment in the very early Universe \citep[Pop III stars, e.g.][]{wha13}. 

While SLSNe are picked out by simple measurement of the total luminosity \citep{gal12},  there is a large
diversity in their spectral types and light curve evolution. By volume, they are rare \citep{qui13} and 
most have been related to massive stars of some sort. However, whether or not their rarity points to 
an origin at the top end of the mass function, or to some other characteristics of the progenitor
(e.g. extremely low metallicity), is not known.
Some SLSNe show clear spectral signatures of hydrogen -- usually relatively narrow emission lines, leading to the classification as type IIn events \citep[e.g.][]{smi07a}. In only one case (SN~2008es) H shows very  broad emission features \citep{gez09, mil09}. Finally,  there appears to be a class of SLSNe that do not show H, and spectroscopically evolve to resemble stripped-envelope type Ic SNe 
\citep[e.g.][]{qui11,pas10b}. These have been termed SLSNe-I by \citet{gal12} and
super-luminous type Ic SNe by \cite{ins13}. 
Originally classified as SLSNe by the first secure redshift determinations, 
they were shown to evolve spectroscopically into type Ic-like SNe on very long timescales by \cite{pas10b}. They were 
discovered up to redshifts $\sim 1$ in the deep Pan-STARRS1 Medium Deep Survey
\citep[][]{cho11} and a significant effort has been invested to characterize them retrospectively
in ongoing surveys \citep{bar09,lel12}. \citet{coo12} and \citet{how13} extended the discovery of SLSNe up to redshift $> 3.9$.

It has been claimed that some of these SLSNe are powered by the radioactive decay of large amounts of \/$^{56}$Ni, and are therefore the observational counterparts of the long sought pair-instability SNe. 
These have even been proposed as a class named ``SLSN-R'' \citep{gal12}. 
The first example is the luminous and slow declining SN~2007bi \citep{gal09}, although \cite{you10} suggested that a standard iron-core-collapse origin could not be ruled out \citep[see also][]{mor13a}; while \citet{dessart12} suggested that energy input from the spin-down of a newly-formed magnetar powers the luminosity, instead of radioactive decay.
Pair-instability SNe are predicted by stellar evolution theory to terminate the lives of stars with mass $>140$ \M\ which have low enough mass-loss rates to retain a large CO-core. 
However, their existence in the local Universe is not settled, since the observed properties of SLSNe can be explained by different physical processes, including magnetar spin-down \citep{woo10,kas10,dessart12,ins13,nic13}; the accretion onto a proto-neutron star or a black hole \citep{dex13}; or interaction of the ejecta with circumstellar material \citep{bli10,che11,gin12,mor13a,mor13b,mor13c}. 

The interaction of the ejecta with CSM is seemingly a frequent phenomenon in massive star explosions, with a number of luminous transients already related to the violent shocks formed when multiple shells collide. 
This is the case for other kinds of outbursts that can compete with SNe in luminosities and expansion velocities \citep[SN impostors,][]{van00,mau06,smi11} even if the star survives the event.
Eruptive mass loss can build up a dense circumstellar medium (CSM) around a massive star. 
When the star eventually explodes, the high-velocity ejecta impacts the dense CSM, converting part of its kinetic energy into radiation. The transient becomes very luminous, with a luminosity evolution modulated by the density profile of the CSM.
When this mechanism is at work it can easily outshine all other radiation sources -- making it difficult, if not impossible, to identify the explosion mechanism or even to assess whether an underlying SN has occurred, as in the spectacular case of SN~2009ip \citep[][and references therein]{pas13,fra13,mau13,mar13}. 

This potential for extreme brightness makes ejecta-CSM interaction an appealing 
physical scenario to  explain supernovae with unusually high luminosities  \citep{che11,gin12}. 
This interpretation has been adopted for H-rich SLSNe, such as SN 2008es \citep{gez09, mil09}, and may also be relevent to H-deprived transients \citep[see e.g.][and references therein]{cha13}. In fact very recently \citet{ben13} presented a case of a Type Ic supernova whose ejecta started to interact with a massive, hydrogen-free ($\sim 3$\M) CSM.

In this paper we report on observations of a new event that appears to support the ejecta--CSM interpretation. CSS121015:004244+132827 (CSS121015 from here on) was first detected by  CRTS on Sep. 15, 2012, at a V magnitude of $\sim 20.0$, as announced by \cite{dra12}. They also reported that a spectrum taken with Palomar 5m Telescope + DBSP on Oct 15th UT showed a very blue continuum with no clear emission features,  resembling the early spectra of other luminous type-I supernovae detected by CRTS. A follow-up spectroscopic observation by \citet{tom12}, obtained with the Asiago 1.82-m Copernico Telescope + AFOSC, confirmed the very blue, featureless continuum, but also allowed the detection of faint, narrow Balmer emissions from the system. A redshift of $\sim 0.286$  was hence derived, making CSS121015 extremely luminous ($\sim -22.5$). 
The spectral appearance was that of a stripped envelope SN such as those discussed in \citet{qui11} and \citet{pas10b}.  
After this classification, we began an extensive follow-up campaign for CSS121015 in the optical domain, within the frameworks of the Asiago-TNG Supernova follow-up campaign\footnote{http://sngroup.oapd.inaf.it} and the Public ESO Spectroscopic Survey of Transient Objects (PESSTO)\footnote{http://www.pessto.org/pessto/index.py}. In this paper we present the results of this monitoring campaign and discuss the implications for the SLSNe scenarios.

\section{Reddening and distance of CSS121015 and its host galaxy} \label{red}

The host of CSS121015 is a faint galaxy, barely visible in Sloan images (SDSS DR9\footnote{http://skyserver.sdss3.org/dr9/en/tools/chart/}) \citep{eis11}. Using our deeper $r$-band frame (see Table \ref{sdss_tab}), we measured an apparent magnitude of $r \sim 23.0$ for the parent galaxy. 

The Galactic reddening towards CSS121015, $A_B$=0.314, was taken from the recalibrated infrared-based dust map of \cite{sf11}. We assume that the extinction in the host galaxy is negligible, as we detect no sign of narrow interstellar NaID absorption \citep{tbc03, poz12}, even in spectra of higher S/N ratio (cf. Section \ref{spec}).

From the narrow emission lines visible in the spectra with better S/N ratio (\Ha, \Hb~and [O~III] 5007\AA~lines; see Section \ref{spec}), we derive a mean redshift $z = 0.2868\pm 0.0006$, where the error is the standard deviation of several measurements.
Assuming a Planck Universe \citep[H$_0=67.3$ km~s$^{-1}$~Mpc$^{-1}$, ${\Omega}_{\lambda} = 0.685$ and ${\Omega}_{m}$ = 0.315 from][]{ade13}, the luminosity distance for CSS121015 is $d_l=1520$ Mpc and the distance modulus is $\mu = 40.91$ mag; these values are used throughout the paper.

Adopting the  reddening and the distance modulus discussed above, we infer a magnitude of $M_r\ga -17.9$ for the faint host galaxy of CSS121015. 

\section{Optical Photometry} \label{phot}
Optical imaging data of CSS121015 were obtained using a number of different instrumental configurations. Basic information on these observations is reported in Tables \ref{obs_tab} and \ref{sdss_tab}. The measured position of the SN in several frames is R.A.$=00^h$42$^m$44$^s.38\pm 0.03$; Decl.$=13^o$28'26$".197\pm 0.05$.

\begin{figure}
\includegraphics[width=8.5cm,angle=0]{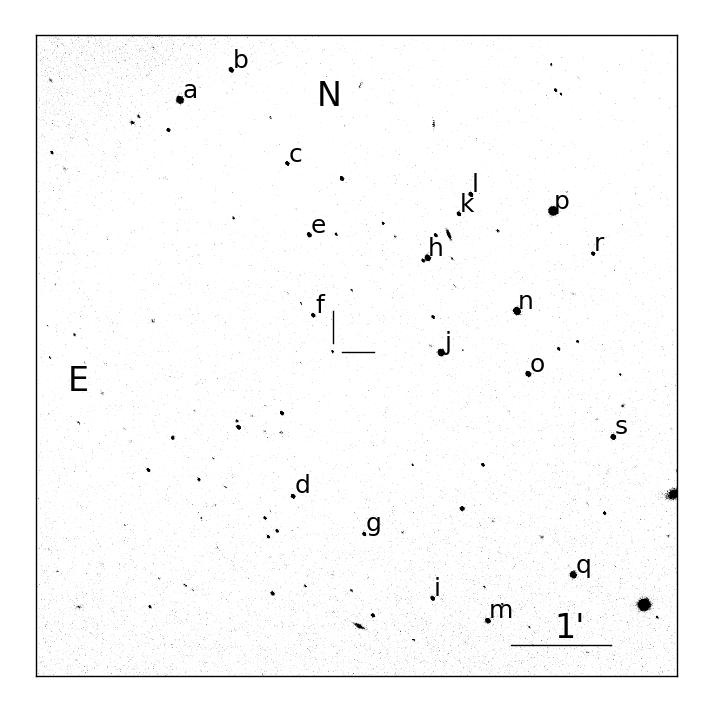}
\caption{CSS121015 supernova and reference stars. This is a TNG+LRS R frame taken on December 25th, 2012. The measured seeing was 1.5".} \label{sn}
\end{figure}

The CCD frames were first bias and flat-field corrected in the usual manner. Since some of the data were obtained under non-photometric conditions, we measured relative photometry with respect to local field stars (see Fig. \ref{sn}), whose magnitudes were computed by averaging estimates obtained on photometric nights.

\begin{table*}
\caption{Magnitudes of local sequence stars identified in Figure~1.}\label{seq}
\begin{tabular}{ccccccccccccc}
\hline
star & RA              & Decl               & U            &errU     &B             & errB     &V             & errV     & R            & errR     & I             & errI    \\
\hline   
a  &  0:42:52.689  &  13:31:46.32  &  15.72  &  0.04  &  15.73  &  0.01  &  15.03  &  0.01  &  14.65  &  0.01  &  14.25  &  0.01 \\
b  &  0:42:49.903  &  13:32:10.22  &  17.87  &  0.02  &  18.06  &  0.02  &  17.32  &  0.01  &  16.95  &  0.01  &  16.52  &  0.04 \\
c  &  0:42:46.848  &  13:30:55.87  &               &             &  19.09  &  0.02  &  18.11  &  0.01  &  17.56  &  0.01  &  17.01  &  0.04 \\
d  &  0:42:46.531  &  13:26:31.35  &               &             &  19.28  &  0.01  &  17.95  &  0.01  &  17.18  &  0.01  &  16.50  &  0.01 \\
e  &  0:42:45.662  &  13:29:59.14  &               &             &  19.80  &  0.01  &  18.12  &  0.01  &  17.12  &  0.01  &  16.01  &  0.02 \\
f  &  0:42:45.450  &  13:28:55.00  &               &             &  20.44  &  0.05  &  18.87  &  0.01  &  17.87  &  0.01  &  16.64  &  0.01 \\
g  &  0:42:42.641  &  13:26:01.14  &  19.10  &  0.08  &  19.30  &  0.02  &  18.53  &  0.01  &  18.14  &  0.01  &  17.70  &  0.02 \\
h  &  0:42:39.203   &  13:29:40.90  &               &             &  17.01  &  0.01  &  16.15  &  0.01  &               &             &               &            \\
i  &  0:42:38.896  &  13:25:09.59  &  17.93  &  0.01  &  18.67  &  0.02  &  17.84  &  0.01  &  17.40  &  0.01  &  16.95  &  0.03 \\
j  &  0:42:38.450  &  13:28:25.51  &  16.27  &  0.02  &  16.11  &  0.01  &  15.31  &  0.01  &  14.89  &  0.01  &  14.50  &  0.02 \\
k  &  0:42:37.487  &  13:30:15.87  &  18.80  &  0.05  &  18.78  &  0.01  &  17.90  &  0.01  &  17.46  &  0.01  &  16.94  &  0.02 \\
l  &  0:42:36.836  &  13:30:31.45  &               &             &  18.53  &  0.02  &  17.53  &  0.01  &  17.01  &  0.01  &  16.48  &  0.03 \\
m & 0:42:35.875  &  13:24:51.53  &  17.16  &  0.04  &  17.45  &  0.02  &  16.82  &  0.01  &  16.43  &  0.01  &  15.95  &  0.02 \\
n  &  0:42:34.314   &  13:28:58.59  &  16.90  &  0.01  &  16.09  &  0.01  &  15.05  &  0.01  &  14.48  &  0.01  &  13.98  &  0.01 \\
o  &  0:42:33.686   &  13:28:08.43  &  17.47  &  0.01  &  17.39  &  0.01  &  16.62  &  0.01  &  16.21  &  0.01  &  15.74  &  0.02 \\
p  &  0:42:32.321  &  13:30:18.29  &  15.56  &  0.01  &  15.10  &  0.01  &  14.14  &  0.01  &  13.65  &  0.01  &               &            \\
q  &  0:42:31.195  &  13:25:28.07  &  17.14  &  0.05  &  16.66  &  0.01  &  15.70  &  0.01  &  15.20  &  0.01  &              &            \\
r   &  0:42:30.116  &  13:29:44.29  &              &             &  19.53   &  0.02  &  18.48  &  0.01  &  17.89  &  0.01  &  17.26  &  0.02 \\
s  &  0:42:29.068  &  13:27:18.36  &  17.71  &  0.03  &  17.69  &  0.01  &  16.80  &  0.01  &               &             &  15.78  &  0.01 \\
\hline
\end{tabular}
\end{table*}

Ideally, one would like to remove the galaxy background by subtraction of a galaxy ``template'' (i.e. an image where the SN is absent). In the case of CSS121015, suitable template images were not available for the Johnson-Cousins frames. However, as the host galaxy is quite faint (see Sect. \ref{red}), the SN magnitudes could be safely measured via PSF-fitting, provided we have decent seeing and the detector pixel scale allows for good PSF sampling. For the actual measurement we used a dedicated, custom-built pipelines (developed by us, E.C.), consisting of a collection of {\sc python} scripts calling standard {\sc iraf} tasks (through {\sc pyraf}), and specific data analysis tools, such as {\sc sextractor} (for source extraction and classification) and {\sc daophot} (for PSF-fitting). In our implementation of the PSF-fitting procedure, an initial estimate of the sky background at the SN location is obtained using a low order polynomial fit to the surrounding region. This is subtracted from the image, before fitting the SN with a PSF model derived from isolated field stars. The fitted source is removed from the original image, an improved estimate of the local background is derived, and the PSF-fitting procedure is iterated. The residuals are visually inspected to validate the fit.

Error estimates were obtained through artificial star experiments, in which a fake star, of similar magnitude to the SN, is placed in the PSF-subtracted residual image in a position close to, but not coincident with that of the real source. The simulated image is processed through the PSF-fitting procedure. This process is repeated at a number of positions; the dispersion of the magnitude returned by  the pipeline is taken as an estimate of the instrumental magnitude error, accounting mainly for the background fitting uncertainty. This is combined (in quadrature) with the PSF-fit error returned by {\sc daophot}, and the propagated errors from the photometric calibration.


\begin{table*}
\caption{Johnson-Cousins photometric measurements for CSS121015, calibrated in the Vega system.}\label{obs_tab}
\begin{flushleft}
\begin{tabular}{crrccccccccccl}
\hline   
 date          &  JD            & phase$^*$    &  U        &errU      & B       &errB      & V         &errV      &R          &errR     &I           &errI        &instrument \\
                   & ($-2400000$) & (days)       &           &             &            &            &            &             &             &            &            &             & \\
\hline   
20120916  &  56186.74  &   -39(-30)	&                   &            &             &           &  20.95  &  0.56   &             &            &            &            &  CRTS \\
20120925  &  56195.87  &	   -30(-23)	&               &            &             &           &  19.76  &  0.28   &             &            &            &            &  CRTS \\
20121006  &  56206.72  &   -19(-15)	&            &            &             &           &  18.82  &  0.20   &             &            &            &            &  CRTS \\
20121015  &  56215.74  &   -10(-8)	&           &            &             &           &  18.55  &  0.35   &             &            &            &            &  CRTS \\
20121019  & 56219.92  &    -6(-5)	&           &            & 18.53   & 0.03   & 18.42   & 0.02    & 18.22   & 0.02    &            &            & AF \\
20121021  &  56221.92  &    -4(-3)	&           &            &             &           &  18.38  &  0.29   &             &           &             &             &  CRTS \\
20121022 & 56222.83    &    -3(-2)	&  17.83 & 0.04 & 18.54 & 0.02 & 18.35 & 0.03 & 18.24 & 0.02 & 18.13 & 0.02 & AF \\
20121026 & 56226.11   &     0(0)	&            &         & 18.29 & 0.20 & 18.19 & 0.10 & 18.20 & 0.06 & 18.00 & 0.07 & RAT \\
20121105 & 56236.99   &     +11(+9)	&               18.17 & 0.04 & 18.70 & 0.02 & 18.52 & 0.02 & 18.33 & 0.02 & 17.91 & 0.02 & AF \\
20121106 & 56238.10   &     +12(+9)	&            18.39 & 0.08 & 18.99 & 0.02 & 18.67 & 0.02 & 18.37 & 0.18 & 18.26 & 0.04 & EF2 \\
20121107 & 56238.80   &     +13(+10)	&           18.23 & 0.03 & 18.85 & 0.02 & 18.45 & 0.02 & 18.30 & 0.02 & 18.25 & 0.02 & AF \\
20121112   &  56243.64 &     +18(+14)	&          &            &             &           &  18.72   &  0.27  &             &          &               &           &  CRTS \\
20121112 & 56244.10    &     +18(+14)	&           18.64 & 0.09 & 19.10 & 0.07 & 18.73 & 0.05 & 18.49 & 0.12 & 18.19 & 0.07 & EF2 \\
20121118 & 56249.05 &        +23(+18)	&           18.62 & 0.02 & 19.22 & 0.02 & 18.72 &  0.02       & 18.47 & 0.02 & 18.15 & 0.02 & LRS \\
20121120  &  56251.59  &     +26(+20)	&             &             &             &           &  18.96  &  0.21   &             &          &               &           &  CRTS \\
20121123 & 56254.13 &        +28(+22)	&           19.17 & 0.09 & 19.52 & 0.04 & 18.98 & 0.02 & 18.75 & 0.02 & 18.34 & 0.03 & EF2 \\
20121130 & 56261.84$^{\ddagger}$ &        +36(+28)	&           19.76 & 0.17 & 19.81 & 0.06 & 19.07 & 0.02 & 18.76 & 0.02 & 18.27 & 0.02 & RAT \\
20121204 & 56265.84 &        +40(+31)	&           19.77 & 0.17 & 19.98 & 0.06 & 19.16 & 0.06 & 18.90 & 0.04 & 18.43 & 0.02 & RAT \\
20121205 & 56266.74 &        +41(+31)      &               &     &         &          &19.34& 0.22 &             &          &               &           &  CRTS \\
20121205 & 56266.85 &        +41(+32)	&            19.85 & 0.16 & 19.92 & 0.04 & 19.22 & 0.03 & 18.81 & 0.02 & 18.39 & 0.02 & RAT \\
20121205 & 56267.00 &        +41(+32)	&              19.90 & 0.09 & 20.03 & 0.04 & 19.32 & 0.05 & 18.96 & 0.04 & 18.46 & 0.04 & EF2 \\
20121206 & 56267.76 &        +42(+33)	&             19.75 & 0.22 & 20.11 & 0.13 & 19.28 & 0.06 & 18.89 & 0.03 & 18.58 & 0.05 & AF \\
20121207 & 56268.85 &        +43(+33)	&            &         & 20.23 & 0.04 & 19.23 & 0.04 & 18.96 & 0.03 & 18.40 & 0.03 & RAT \\
20121209 & 56270.85 &        +45(+35)	&            &         & 20.24 & 0.06 & 19.34 & 0.03 & 19.05 & 0.02 & 18.44 & 0.02 & RAT \\
20121212 & 56274.10 &       +48(+37)	&            20.31 & 0.10 & 20.43 & 0.04 & 19.57 & 0.02 & 19.13 & 0.15 & 18.61 & 0.05 & EF2 \\
20121215 & 56276.84 &       +51(+40)	&      &         & 20.55 & 0.11 & 19.56 & 0.04 & 19.06 & 0.03 & 18.53 & 0.03 & RAT \\
20121217 & 56278.81 &       +53(+41)	&      &         &              &         & 19.74 & 0.04 & 19.17 & 0.04 &              &         & AF \\
20121218 & 56279.87 &       +54(+42)	&      &         & 20.67 & 0.10 & 19.58 & 0.03 & 19.14 & 0.02 & 18.68 & 0.02 & RAT \\
20121218 & 56279.88 &       +54(+42)	&      &         & 20.68 & 0.13 &              &         &              &         &              &         & AF \\
20121222 & 56284.00 &       +58(+45)	&            20.64 & 0.10 & 20.96 & 0.11 & 19.85 & 0.03 & 19.26 & 0.03 & 18.72 & 0.02 & LRS \\
20121225 & 56286.75 &       +61(+47)	&            20.68 & 0.09 & 20.97 & 0.05 & 19.94 & 0.03 & 19.40 & 0.02 & 18.93 & 0.02 & LRS \\
20121227 & 56288.85 &       +63(+49)	&      &         & 20.99 & 0.14 & 19.95 & 0.11 & 19.47 & 0.05 & 18.76 & 0.02 & RAT \\
20130102 & 56295.10 &       +69(+54)	&      &         & 21.43 & 0.04 & 20.26 & 0.03 & 19.64 & 0.03 & 18.91 & 0.05 & EF2 \\
20130108 & 56300.75 &       +75(+58)	&             21.65 & 0.06 & 21.60 & 0.02 & 20.50 &   0.02      & 19.69 & 0.02 & 19.00 & 0.02 & LRS \\
20130112 & 56304.83 &       +79(+61)&      &         & 21.90 & 0.06 & 20.56 & 0.05 & 19.86 & 0.03 & 19.10 & 0.06 & RAT \\
20130115 & 56307.88 &       +82(+64)&              &         & 22.12 & 0.10 & 20.65 & 0.18 & 20.05 & 0.07 & 19.20 & 0.03 & RAT \\
20130124 & 56316.87 &       +91(+71)&       &         & 22.24 & 0.26 & 21.02 & 0.15 & 20.25 & 0.06 & 19.43 & 0.02 & RAT \\
20130127 & 56320.00 &       +94(+73)&       &         &              &         & 21.27 & 0.11 & 20.18 & 0.07 & 19.56 & 0.17 & LRS \\
20130128 & 56320.84 &       +95(+74)&      &         & 22.58 & 0.18 & 21.08 & 0.13 & 20.29 & 0.06 & 19.54 & 0.04 & RAT \\
20130130 & 56322.74 &       +97(+75)&       &         &              &         & 21.21 & 0.14 & 20.37 & 0.06 & 19.71 & 0.08 & AF \\
\hline
\end{tabular}

* - Relative to the estimated epoch of the V maximum (JD = 2456226); the phase in parenthesis is in the SN rest frame. The rise to maximum V lasted $\sim 51$ days ($\sim 40$ days in the rest frame).\\
CRTS = Catalina Real-time Transient Survey \\
AF = Asiago 1.82m  Telescope + AFOSC\\
RAT = Liverpool Telescope + RATCam (SDSS u, r, i and Johnson B, V)\\
EF2 = ESO-NTT + EFOSC2\\
LRS = TNG + LRS \\
$\ddagger$ for JD=2456264.72 we have also NIR LBT-Lucifer observations (on 2MASS scale): J=$18.46\pm 0.05$; H=$18.34\pm 0.04$; K=$17.94\pm 0.06$.\\

\end{flushleft}
\end{table*}

\subsection{Johnson-Cousins UBVRI photometry}\label{lc}

\begin{figure}
\includegraphics[width=8.5cm,angle=0]{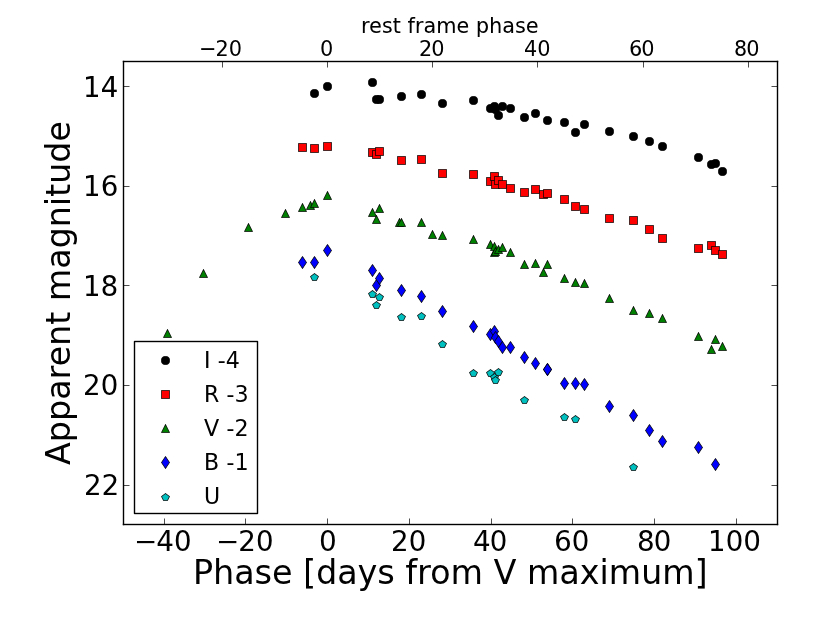}
\caption{Johnson-Cousins U (top),B,V,R and I (bottom) light curves of CSS121015. The phase given is in the observer frame (bottom x-axis) and SN rest-frame (top x-axis).}
\label{phot_fig}
\end{figure}

Six photometric nights were used to calibrate the local stellar sequence, introduced in Section \ref{phot}, against Landolt standard stars
\citep{land}. The Johnson-Cousins magnitudes of the local standards, and their estimated errors, are shown in Table \ref{seq}.
In addition to standard UBVRI filters, CSS121015 was also observed with SDSS filters (cfr. Sect.~\ref{lc_sdss}).
In particular, we used the Liverpool Telescope (+ RATCam) to get SDSS-{\it u}, Johnson-BV and SDSS-{\it ri}. These intrumental magnitudes were then transformed to Johnson-UBV and Cousins-RI magnitudes using colour equations, again derived through observations of Landolt fields, to enrich the UBVRI light curves.  The final SN magnitudes (calibrated in the Vega system) are presented in Table~\ref{obs_tab}.  Sometimes the errors are relatively high, mostly because of poor sky transparency and/or poor seeing. The UBVRI light curves are plotted in Figure \ref{phot_fig}. 

Only the V-band light curve has pre-maximum points. They are from unfiltered (pseudo-V) CRTS observations. CRTS magnitudes have been obtained after template subtraction (see Sect. \ref{lc_sdss} for a description of this technique), with the template taken on June 22nd, 2012, well before the SN explosion. The magnitudes have been scaled to V using our local standards (see Table \ref{seq}). The rising phase is steeper than the post-maximum decline. By fitting a low order polynomial to the observed V-band light curve, we estimate that the maximum occurred on JD = $2456226\pm2$, at V$ =18.34\pm 0.10$ mag. Given the smooth rise shown by the V light curve, we fit a parabola to the flux derived from the first three points and estimate that the explosion occurred on JD = $2456175\pm 10$, e.g. about ten days before discovery and 40 days before V-band maximum light (rest frame). The uncertainty in the explosion epoch accounts for the possibility of an early ``plateau''-like break, as seen in the rising light curve of one SLSN \citep[see][]{lel12}. Although the explosion epoch would allow the most physically meaningful comparisons with other SNe, its large associated error means that we prefer to use the time of maximum light as a reference epoch.


The post-maximum declines are approximately linear in all bands, and the light curves never show a break in their decline. The observed declines are progressively slower as we move from the U to the I band. Starting from maximum light, the observed linear decline rates are $5.40\pm0.20$, $4.53\pm0.10$, $3.23\pm0.08$, $2.43\pm0.08$, and $1.92\pm0.11$ mag $(100 \mathrm{d})^{-1}$ in U, B, V, R and I, respectively.

\subsection{SDSS $griz$ photometry}\label{lc_sdss}

CSS121015 was observed in the SDSS {\it griz} bands with the Faulkes North Telescope, equipped with the MEROPE1 and Spectral2 cameras. Table~\ref{sdss_tab} integrates this data with the {\it ri} data from LT+RATCam, mentioned above, collecting all SDSS photometry. This was reduced with the same recipes used for the Johnson-Cousins photometry. The photometric calibration was achieved by comparing the magnitudes obtained for the stars in the field of CSS121015 with their SDSS magnitudes, and is therefore close to the AB system ($g,r,i_\mathrm{SDSS} \sim g,r,i_\mathrm{AB}$, while $z_\mathrm{SDSS} \sim z_\mathrm{AB} - 0.02$ mag\footnote{http://www.sdss3.org/dr9/algorithms/fluxcal.php\#SDSStoAB}; all magnitudes given in Table \ref{sdss_tab} are in the ABmag system.).

In the table we also report deep {\it r} and {\it i} limits obtained with WHT+ACAM and TNG+LRS at about 200 days (rest frame) after maximum.
These will turn out to be very important in deriving an upper limit to the $^{56}$Ni mass synthesized in the explosion (see Sect. \ref{disc}). \\
Given the low surface brightness of the host galaxy, we were able to obtain reliable deep limits directly on the {\it r} and {\it i} images. However, we checked these with template subtraction. This technique requires that exposures of the field, obtained before the explosion or long after the SN has faded, are available. The templates have be taken with the same filters used for SN imaging, with high S/N ratio and good seeing. While in principle they should be obtained with the same telescopes as those used for the specific SN observations (to guarantee the same bandpass), in practice archival images with the proper filters in the on-line archives are very welcome. In our case we were able to retrieve deep pre-discovery exposures in the $r,i$ bands from SDSS3-DR9. 

In template subtraction, the template image is first geometrically registered to the same pixel grid as the SN frame. The PSF of the two images is then matched by means of a convolution kernel determined by comparing a number of reference sources in the field (for this we used {\sc hotpants}\footnote{http://www.astro.washington.edu/users/becker/hotpants.html}). After matching the photometric scale, the template image is subtracted from the SN image, and a difference image is obtained. In our case no residuals were detectable in the $r-$ and $i-$band frames at the SN location, so we can only estimate an upper limit to the SN magnitude.

 The limits were estimated through artificial star experiments, in which a fake star of a given magnitude is placed in the original frame at the precise position of the source. The process is iterated several times by injecting fainter and fainter fake stars. The limit is given by the magnitude of the fake star that leaves detectable residuals after template subtraction. As already stated, the limits obtained with template subtraction agree well with those derived directly on the deep observed images (given in Table \ref{sdss_tab}).

\begin{table*}
\caption{SDSS Photometry in {\it griz} bands, calibrated in the ABmag system.}\label{sdss_tab}
\begin{flushleft}
\begin{tabular}{ccrccccccccl}
\hline   
 date          &  JD            & phase$^*$    &  {\it g}   &err{\it g}    & {\it r}     &err{\it r}     & {\it i}     &err{\it i}     &{\it z}     &err{\it z}   &instr$^{**}$ \\
                  & ($-2400000$) &  (days)           &             &                &               &               &              &                &             &                &              \\
\hline
20121026 & 56226.10 & 0(0)	&         &    & 18.42 & 0.12 & 18.25 & 0.18 &         &    & RAT \\
20121030 & 56231.00 & +5(+4)	&          18.50 & 0.30 & 18.42 & 0.16 &         &    &         &    & Fau2 \\
20121102 & 56233.90 &  +8(+6)	&         18.51 & 0.04 & 18.49 & 0.02 & 18.62 & 0.07 & 18.56 & 0.06 & Fau2 \\
20121105 & 56236.70 &  +11(+9)	&         18.62 & 0.04 & 18.52 & 0.02 & 18.58 & 0.07 & 18.59 & 0.04 & Fau2 \\
20121109 & 56240.70 &  +15(+12)	&        18.69 & 0.06 & 18.57 & 0.02 & 18.65 & 0.07 & 18.55 & 0.04 & Fau2 \\
20121111 & 56242.70 &   +17(+13)	&         18.83 & 0.05 & 18.63 & 0.04 & 18.67 & 0.09 & 18.56 & 0.08 & Fau2 \\
20111115 & 56246.80 &   +21(+16)	&          18.84 & 0.07 & 18.73 & 0.04 & 18.75 & 0.08 &         &    & Fau2 \\
20121126 & 56257.80 &   +32(+25)	&         19.18 & 0.09 & 18.78 & 0.13 & 18.76 & 0.10 & 18.83 & 0.10 & Fau1 \\
20121130 & 56261.84 &   +36(+28)	&                   &    & 18.97 & 0.04 & 18.75 & 0.03 &         &    & RAT \\
20121202 & 56263.70 &   +38(+30)	&         19.51 & 0.06 & 19.02 & 0.04 & 18.94 & 0.07 & 18.67 & 0.10 & Fau1 \\
20121204 & 56265.83 &   +40(+31)	&                   &    & 19.10 & 0.03 & 18.91 & 0.03 &         &    & RAT \\
20121205 & 56266.84 &   +41(+32)	&                   &    & 19.04 & 0.03 & 18.85 & 0.03 &         &    & RAT \\
20121207 & 56268.84 &   +43(+33)	&                   &    & 19.21 & 0.04 & 18.91 & 0.06 &         &    & RAT \\
20121209 & 56270.84 &   +45(+35)	&                   &    & 19.29 & 0.03 & 18.96 & 0.03 &         &    & RAT \\
20121214 & 56275.80 &   +50(+39)	&        19.96 & 0.08 & 19.31 & 0.08 & 19.03 & 0.09 &         &    & Fau1\\
20121215 & 56276.83 &   +51(+40)	&                  &    & 19.34 & 0.04 & 19.00 & 0.05 &         &    & RAT \\
20121218 & 56279.86 &   +54(+42)	&                   &    & 19.39 & 0.03 & 19.11 & 0.03 &         &    & RAT \\
20121227 & 56288.84 &   +63(+49)	&                  &    & 19.74 & 0.05 & 19.29 & 0.03 &         &    & RAT \\
20130112 & 56304.82 &   +79(+61)	&                   &    & 20.19 & 0.05 & 19.67 & 0.12 &         &    & RAT \\
20130115 & 56307.87 &   +82(+64)	&                   &    & 20.34 & 0.06 & 19.90 & 0.08 &         &    & RAT \\
20130124 & 56316.86 &   +91(+71)	&                   &    & 20.61 & 0.07 & 20.00 & 0.04 &         &    & RAT \\
20130128 & 56320.83 &   +95(+74)	&                   &    & 20.65 & 0.07 & 20.10 & 0.09 &         &    & RAT \\
20130610  &56453.75  &  +228(+177)	&                   &    &$>$22.4 & 0.2&            &         &         &    & ACA\\
20130611 & 56454.67  &  +229(+178)	&                    &    &           &        & $>$22.4 & 0.2 &         &    & ACA\\
20130715 & 56488.69&    +263(+204)	&                   &    &$>$23.1 & 0.3&            &         &         &    & LRS\\
\hline
\end{tabular}

* - Relative to the estimated epoch of V maximum light (JD = 2456226); the phase in parenthesis is in the SN rest frame.\\

** - See note to Table \ref{obs_tab} for instrument coding, plus:\\

Fau1 =  Faulkes Telescope North+EM01\\
Fau2 =  Faulkes Telescope North+fs02\\
ACA = WHT + ACAM\\

\end{flushleft}
\end{table*}

\section{Spectroscopy} \label{spec}
Our spectroscopic observations cover a time interval from day -6 to day +85 (in the observer frame) from the estimated maximum epoch.  Table \ref{spec_tab} lists the date (column 1), the JD (col. 2), the observed (and rest-frame) phases relative to the explosion (col. 3), the wavelength range (col. 4), instrument used (col. 5), and the resolution as measured with a gaussian fit of the night-sky lines (col. 6). For some epochs, we co-added near-contemporaneous spectra to improve the S/N. 
\begin{figure*}

\includegraphics[width=16cm,height=21cm,angle=0]{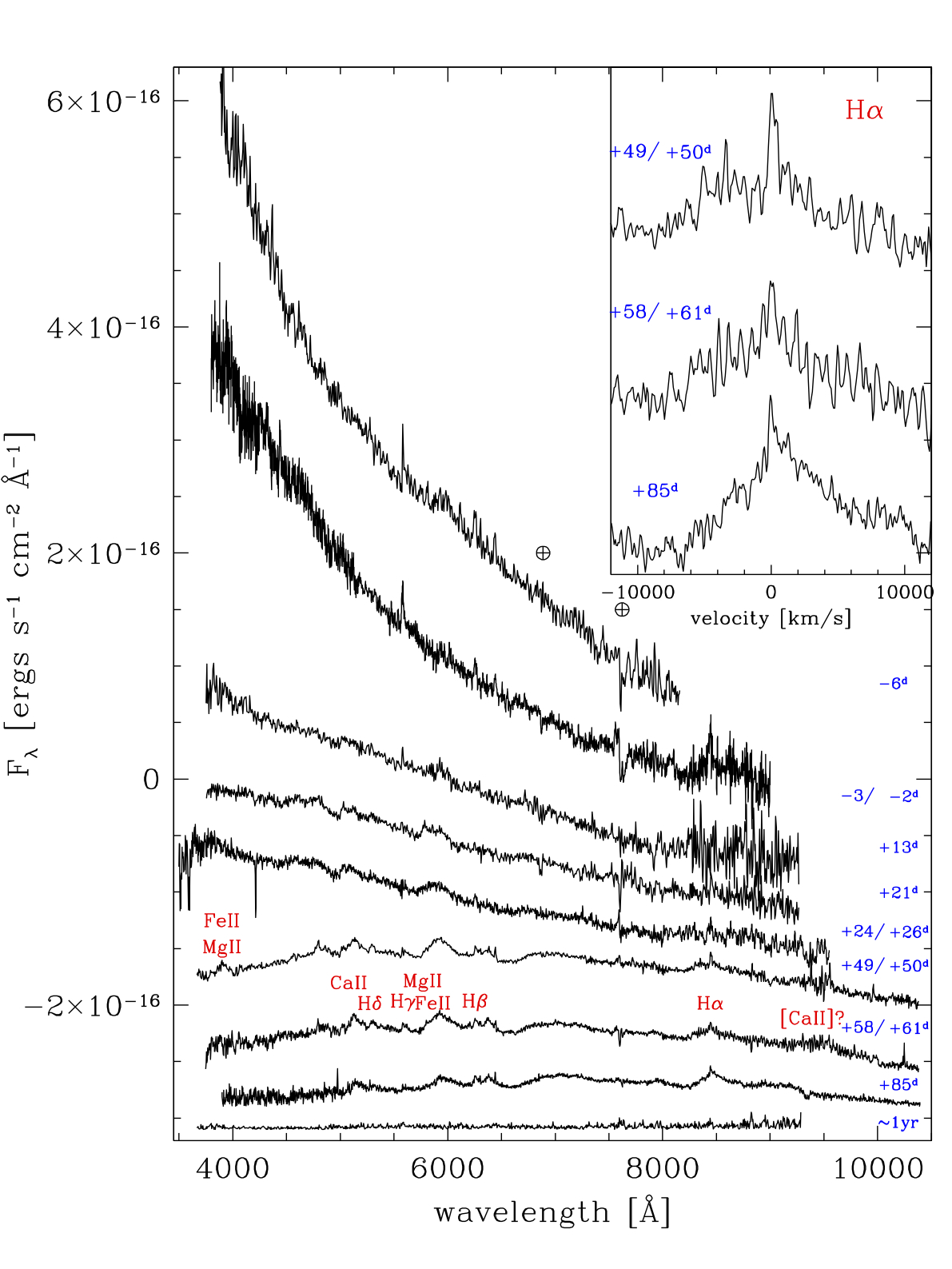}
\caption{Spectral evolution of CSS121015. Wavelength and phases (from maximum) are in the observer's frame. The ordinate refers to the first spectrum, and the others have been arbitrarily shifted downwards. The -3/-2 days spectrum is a merge of the AFOSC and ISIS spectra. The +13 days spectrum is the average of the AFOSC and EFOSC2 spectra. The +24/+26 days spectrum is a merge of the LRS and ISIS spectra. The +49/+50 days spectrum is the average of the EFOSC2 and LRS spectra. Finally, the +58/61 days spectrum is the average of two LRS spectra (see Table \ref{spec_tab}). The last spectrum is the host galaxy spectrum taken about 400 days after maximum. Residuals from the atmospheric absorption corrections have been marked with the $\oplus$~symbol. The inset shows the evolution of the \Ha~profile, in velocity space.}
\label{spec_evol}
\end{figure*}

The spectra were reduced using standard {\sc iraf} routines. Extractions were variance-weighted, based on the data values and a poisson/ccd model using the gain and readout-noise parameters. The background at either side of the SN signal was fitted with a low order polynomial and then subtracted. Fluxing and telluric absorption modeling were achieved using observations of spectrophotometric standard stars. 
For EFOSC spectra, all the previous steps were optimized in a custom-built {\sc python/pyraf} package (PESSTO-NTT pipeline) developed by one of us (S.V.) including also an automatic check of the wavelength calibration using the sky lines.
Most spectra have been taken with the slit aligned along the parallactic angle. The flux calibration of the spectra was checked against the photometry (using the {\sc iraf} task {\sc stsdas.hst\_calib.synphot.calphot}) and, when discrepancies occurred, the spectral fluxes were scaled to match the photometry. On nights with fair to good sky conditions, the agreement with photometry was within 15\%.  A selection of the CSS121015 spectra is shown in Figure \ref{spec_evol}.

The last EFOSC2 spectra, taken about one year after the SN explosion (rest frame) when the SN had faded away, is that of the host galaxy. To increase the S/N, we co-added the six spectra for a total exposure time of 16200 sec. In order to put the faint host galaxy in the slit, we used star f (see Figure \ref{sn}) as guide and rotated the slit to a position angle of 28.3 degree. The photometry synthesized from the spectra of star f (in comparison with the magnitudes given in Table \ref{seq}) was used to normalize the flux of the galaxy spectra. While the host galaxy spectra were obtained under low-to-average sky conditions, after computing a weighed mean of the six individual spectra, according to the S/N, the resulting combined spectrum (see boxes a and b of Figure \ref{spec_evol}) was good enough for our purposes (see Sect. \ref{sp_evol}).

\subsection{Spectroscopic evolution and comparison with other SLSNe-II}\label{sp_evol}

\begin{figure}
\includegraphics[width=9cm,angle=0]{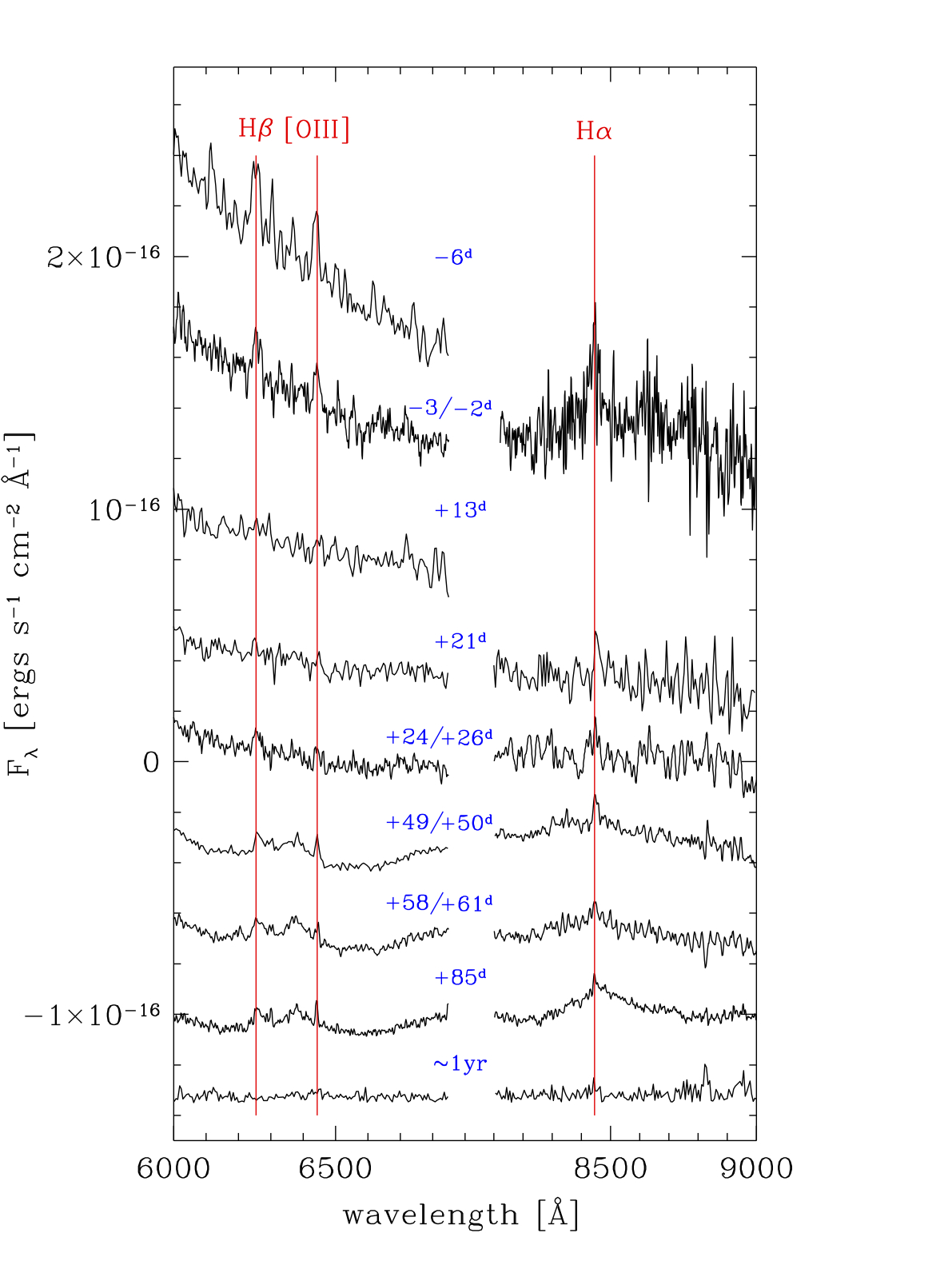}
\caption{Spectral evolution of CSS121015 zoomed on the narrow \Hb, [O~III] 5007\AA~ and \Ha~transitions. Wavelength and phases are as in Figure \ref{spec_evol}.
}
\label{narrow_spec}
\end{figure}

The early spectra show a very blue continuum ($T_{\mathrm{BB}}\sim 17700$\,K, after reddening and redshift corrections). This becomes
progressively redder with phase, reaching a temperature of about $5500$\,K at +85 days. Initially, the spectra are almost featureless, with the first broad features -- mostly  due to Fe II lines -- becoming visible in the spectrum at +21 days. The spectrum at -3/-2 days (see Figs. \ref{spec_evol} and \ref{narrow_spec}) shows a narrow, barely resolved \Ha~emission, and no clear broad Balmer features.

\begin{figure}
\includegraphics[width=9cm,angle=0]{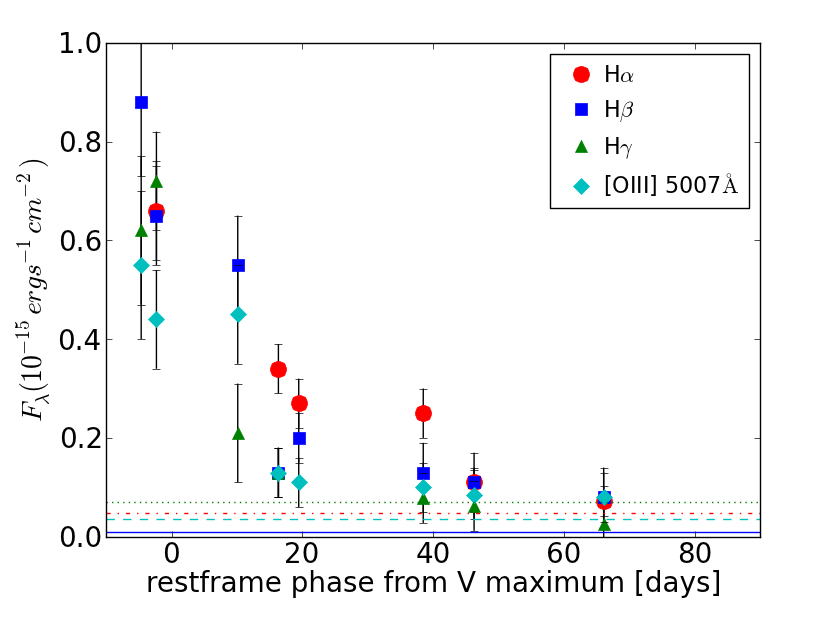}
\caption{Flux evolution of the narrow Balmer and [O~III] 5007\AA~ lines. The reported flux is the true flux of the lines above the continuum. The lines mark the values of the upper limits derived for the four emission lines in the host spectrum taken on November-December 2013 (see Table \ref{spec_tab}; dash-dotted, red, line refers to \Ha~emission; solid, blue, line to \Hb; dotted, green, line to \Hg; and dashed, cyan, line to [O III] 5007\AA). The last deep spectra of the host galaxy show that the contamination of the narrow line emissions from the host is negligible.
}
\label{narrow}
\end{figure}

However, the high-S/N ratio spectra at phases +58/+61 days show a dim and broad (FWMH $\sim10000$ \kms) \Ha~emission, with the faint, narrow component superimposed (and only marginally resolved on top of it). In the blue part of the spectrum, narrow Balmer and [O III] emission lines are also visible. From the narrow Balmer lines of the +85 day spectrum we derive a gaussian FWHM $\la 600$ \kms. Whether the narrow lines are intrinsic to the SN or are of interstellar origin is difficult to establish, mostly because of the relatively low S/N ratio of the first spectra, although the upper limit of 600 \kms~of the narrow Balmer lines is consistent with fast wind of massive stars. 
Moreover, the luminosities of the narrow hydrogen (\Ha,~\Hb,~and \Hg) and O~III $5007$\AA~line show a smooth decrease (see Figure \ref{narrow}), suggesting that the the narrow H and [O III] 5007\AA~lines are related to the SN event. To settle this issue, we analyzed the host galaxy spectrum taken almost a year after the SN went off. Given the faintness of the host, the spectrum is relatively noisy, but it shows a weak continuum from which a synthetic {\it r} magnitude of $22.7 \pm 0.5$ was derived, that is consistent with the magnitude derived from deep imaging (see Table \ref{sdss_tab}) of $r \sim 23.0$. There is no sign of narrow emission line in the co-added spectrum (see Figures \ref{spec_evol}, \ref{narrow_spec} and \ref{narrow}), and we derive the following upper limits: F(H$\gamma)\le 7.0\times 10^{-17} {\rm erg}\,{\rm s}^{-1} {\rm cm}^{-2}$; F(H$\beta)\le 1.0\times 10^{-17} {\rm erg}\,{\rm s}^{-1} {\rm cm}^{-2}$; F([OIII 5007\AA])$\le 3.5\times 10^{-17} {\rm erg}\,{\rm s}^{-1} {\rm cm}^{-2}$; and F(H$\alpha)\le 4.9\times 10^{-17}  {\rm erg}\,{\rm s}^{-1} {\rm cm}^{-2}$. 

These upper limits prove that the narrow lines seen in the early spectra are indeed intrinsic to the SN event and that CSM was surrounding the SN ejecta.
Interestingly, flux variation of [O III] and Balmer lines has previously been detected in SNe~IIn \citep[e.g. in][]{mt88z}.

Given the clear presence of a broad \Ha~line in the spectra later than +49/+50 days and of narrow lines along its spectral evolution, we may finally classify CSS121015 as a Type II/IIn supernova.

At this time, the blue part of the spectrum is dominated by broad metal features, mostly belonging to the Fe group, with some contribution from Ca II H\&K and Mg II.
The spectra show little evolution between 50 days and the end of our observations, and never show any hint of O, though we detect what may be a weak [Ca II] 7300 \AA~line. As shown in Figure \ref{narrow_spec} and in the inset of Figure \ref{spec_evol}, the \Ha~emission shows a slightly asymmetric, triangular profile, with a red wing extending up to $v_{ZI}\sim 10000$ \kms.

\begin{table}
\caption{Spectroscopic observations of CSS121015} \label{spec_tab}
\begin{tabular}{rcrlrr}
\hline
\hline
date   & JD    & phase$^*$  &   range   & inst.$^{**}$      & res.\\
       &$-2400000$        & (days)     &   (\AA)   &                  &(\AA)\\
\hline			    		                    
20121019 &56220.38& -6(-4)& 3500-8200 & AF &  13.5  \\
20121022 &56223.29& -3(-2)& 3500-8200 & AF  &  13.5  \\
20121022& 56223.62& -2(-1.5)&  3200-9000 & ISI &    6       \\
20121105&56237.38& +11(+9)& 3500-8200 & AF   &  13.5  \\
20121107&56238.65& +13(+10)& 3650-9300 & EF2  &  18     \\
20121107&56239.33& +13(+10)& 3500-8200 & AF    &  13.5  \\
20121113&56244.58& +19(+15)& 3650-9300 & EF2  &  18     \\
20121115&56246.56& +21(+16)& 3650-9300 & EF2  &  18     \\
20121118&56249.50& +24(+19)& 3200-8000 & LRS          &  10.5  \\
20121120&56252.47& +26(+20)& 3500-9800 & ISI             &  7    \\
20121123&56254.59& +29(+23)& 3650-9300 & EF2           &  18     \\
20121203&56264.76& +39(+30)& 9500-13500&Luc            &    3   \\
20121204&56266.30& +40(+31)& 3500-8200 & AF             &  13.5  \\
20121206&56267.56& +42(+33)& 3650-9300 & EF2           &  18     \\
20121209&56271.29& +45(+35)& 3500-8200 & AF             &  13.5  \\
20121213&56274.55& +49(+38)& 3650-9300 & EF2           &  18     \\
20121214&56276.46& +50(+39)& 3200-10000 & LRS          &  10.5  \\
20121222&56284.38& +58(+45)& 3200-10000 & LRS          &  10.5  \\
20121225&56287.36& +61(+47)& 3200-10000 & LRS          &  10.5  \\
20130118&56311.33& +85(+66)& 3600-10000 & OSI          &  9  \\
20131126&56622.61&+397(+309)&3650-9250 & EF2         &28 \\
20131202&56628.54&+403(+313)&4800-9250 & EF2         &28 \\
20131203&56629.55&+404(+314)&4800-9250 & EF2         &28 \\
20131225&56651.56&+426(+331)&4800-9250 & EF2         &28 \\
20131226&56652.59&+427(+332)&4800-9250 & EF2         &28 \\
20131227&56653.56&+428(+333)&4800-9250 & EF2         &28 \\

\hline
\end{tabular}

* - Relative to the estimated epoch of the V maximum (JD = 2456226); the phase in parenthesis is in the SN rest frame.

** - See note to Table \ref{obs_tab} for instrument coding, plus:\\
ISI = WHT+ISIS \\
Luc = LBT+Lucifer (flat continuum, noisy spectrum)\\
OSI = GTC+OSIRIS \\

\end{table}

\begin{figure*}
\includegraphics[width=12cm,height=15.0cm,angle=0]{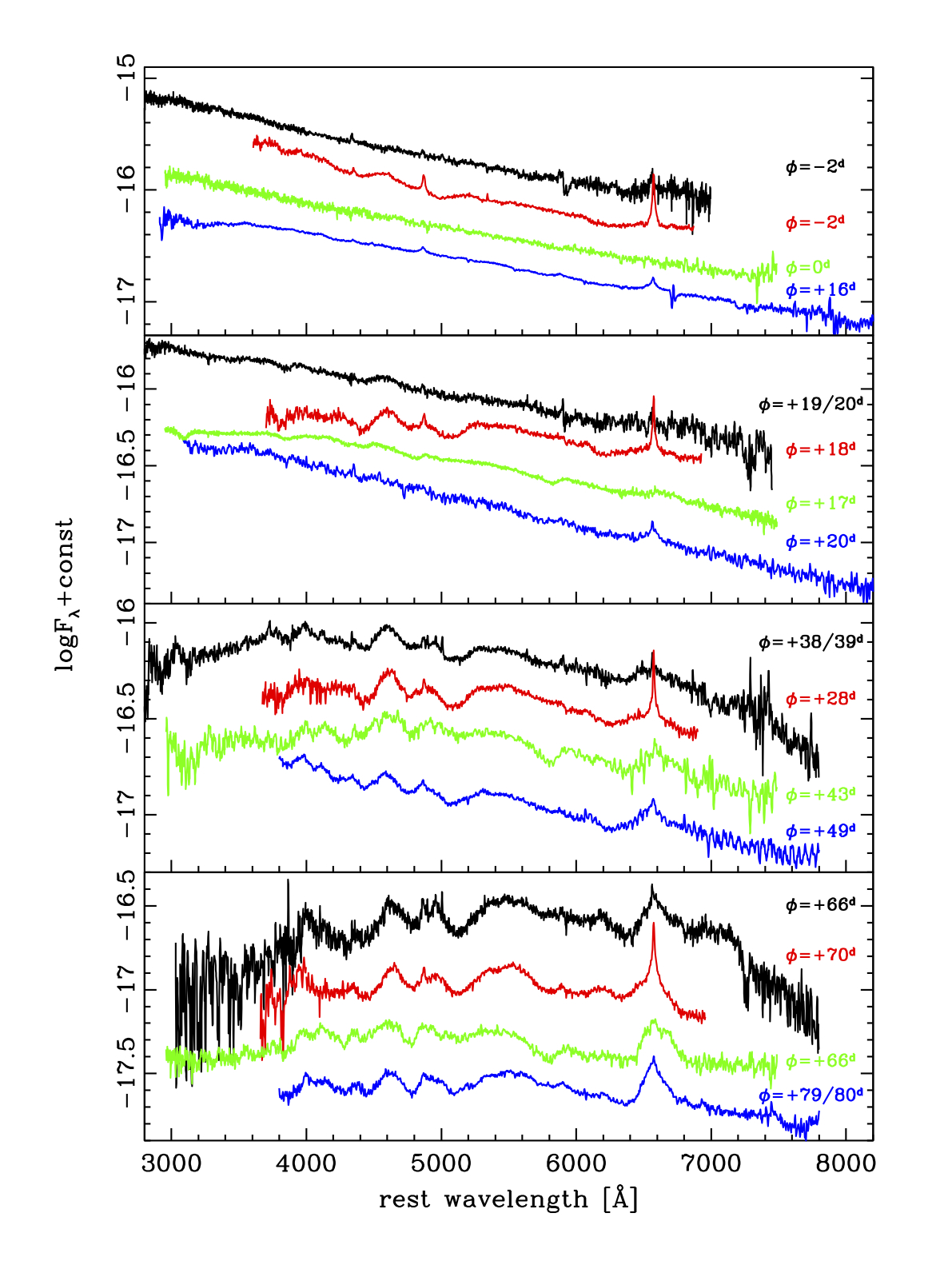}
\caption{Comparison of the spectra of CSS121015 with those of a selected sample of luminous SNe at significant epochs (cfr. Sect.~\ref{sp_evol}).
From top to bottom in each panel: spectra of CSS121015 (black), SN 2005gj \citep[][red]{pri07,ald06}, SN 2008es \citep[][green]{gez09,mil09} and SN 2008fz \citep[][blue]{agn10}. The spectra have been corrected for redshift and reddening. The rest frame phases are from the estimated dates of maximum.
}
\label{spec_conf}
\end{figure*}

In Fig. \ref{spec_conf}, the CSS121015 spectra are compared with other very luminous SNe at similar phases. The comparison SN sample (SNe~2005gj, 2008es, and 2008fz) has been selected using the GELATO comparison tool\footnote{https://gelato.tng.iac.es; the list of GELATO templates is given here: https://gelato.tng.iac.es/templates} \citep{har08} as the objects that give the best overall match to CSS121015 in particular at late times, when the SN features became stronger.
Up to day $\sim +20$ (rest frame), the spectra are dominated by a blue continuum, and only the +18 day SN 2005gj spectrum shows broad features, at this time. In the Ejecta-CSM interaction scenario, this can be attributed to the CSM shell in SN 2005gj having a smaller radius (fainter luminosity, see Figure \ref{Mb_fig}) and lower density than the CSM around the other SNe, therefore revealing the underlying SN earlier. The broad bump at $\sim 4600$ \AA, visible in the +19/20 days CSS121015 spectrum, is also detected in SN 2008es at a similar phase, while the SN 2008fz spectrum at a slightly later phase (+49 days) shows stronger features. At +38/39 days, the spectra of the four SNe are extraordinarily similar, with the exception of the narrower \Ha~components, which are stronger in SN 2005gj. This SN shows prominent intermediate and narrow line profiles, which suggests the presence of a more extended, dilute, and H-rich CSM. The broad components have similar profiles, with terminal velocities of $\sim 10000$ \kms. The spectra at late phases also show very good agreement, and are all dominated in the blue by transitions of iron group elements. The spectral energy distributions (SEDs) of the SNe in our sample are very similar at all epochs

\section{Bolometric light curve}\label{bol}

The pseudo-bolometric light curve of CSS121015 has been computed by integrating its multi-colour photometry from U to $z$, neglecting any possible contribution in the low (infrared/radio) and high energy (UV, X-ray) domains. We proceeded by deriving, for each epoch and filter, the flux at the effective wavelength. We adopted as reference the epochs of the V-band photometry, and missing measurements at given epochs for the other filters were obtained through interpolation or, if necessary, by extrapolation assuming a constant colour from the closest available epoch. The fluxes at the filter effective wavelengths, corrected for extinction, provide the spectral energy distribution at each epoch, which is integrated by the trapezoidal rule, assuming zero flux at the integration boundaries. The observed flux was converted into luminosity for the adopted distance.
The pseudo-bolometric light curve is shown in Figs. \ref{bolom_model} and \ref{bolom_conf}.

Given the high temperature, the wavelength ranges of our optical spectra only sample the Rayleigh-Jeans tail of the CSS121015 SED. This means that the temperatures derived from the early spectra have significant errors (up to $\sim$10\%). This also implies that there is significant emission outside the observed spectral range. If we assume that the SED can be well fitted with a black-body (BB), and we calculate the flux below it, the extrapolated flux that may be regarded as an upper limit to the real flux emitted by CSS121015 in that particular phase, because in real SNe the UV region of the spectra can be affected by severe line-blanketing, that shifts the emitted flux to longer wavelengths. This is why the integrated BB emission is giving us only an upper limit to the real SED. We see evidence for this line-blanketing in the later, cooler spectra, with  SED maxima well inside the observed wavelength range. The true luminosity falls somewhere between the pseudo-bolometric luminosity and the BB extrapolation. Approximating the SN as a BB also allows us to estimate a radius of the emitting region.

The BB extrapolated light curve is shown in Fig. \ref{bolom_model}. The CSS121015 bolometric light curve shows a relatively fast rise to maximum of about 40 days, and a slower post-maximum linear decline up to phase of about 80 days past maximum in the rest frame. It reaches a very bright peak of about log $L$ = 44.50 dex ($\sim 44.9$ in the BB extrapolation, see Table \ref{rad}). The late deep $r,i$ photometric limits (see Table \ref{sdss_tab}) translate into an upper limit of log $L$ = 42.40 dex.

\section{Discussion}\label{disc}

 \subsection{Absolute magnitudes of CSS121015 and comparison with other SLSNe} \label{abs}

\begin{figure*}
\includegraphics[width=15cm,angle=0]{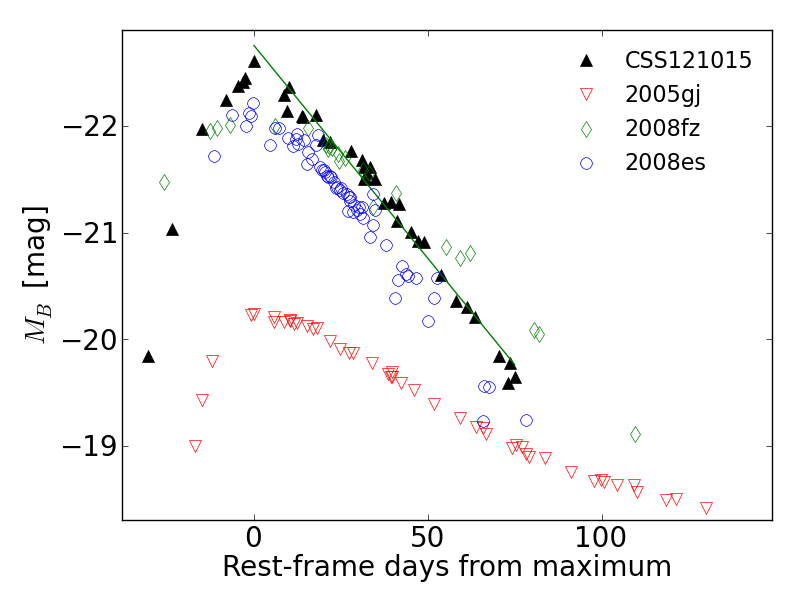}
\caption{Comparison of the absolute B-band light curve of CSS121015 with those of SNe 2005gj, 2008es and 2008fz. The phases have been corrected for the time dilation due to cosmic expansion; a K-correction has been applied, following the \citet{kim96} recipe.
SN 2005gj is the nearest supernova \citep[$z=0.06$ and photometry from][]{pri07, ald06}. We have adopted only Galactic absorption correction \citep[from E(B$-$V) = 0.107 mag,][]{sf11} and maximum epoch JD = 2453658 \citep{pri07}.
SN 2008es is at a distance similar to that of CSS121015 (z = 0.202 taken from the Asiago Supernova Catalogue, ASC).
The photometry and maximum epoch (JD = 2454603) are from \citet{gez09} and Galactic reddening (E(B$-$V) = 0.010 mag) from \citet{sf11}. 
For SN 2008fz (redshift z = 0.133, from ASC) the photometry and maximum epoch (JD = 2454731) are from \citet{dra10}, \citet{agn10} and Benetti et al. (in preparation). A correction for Galactic reddening of E(B-V) = 0.036 mag \citep{sf11} has been applied. The linear fit of CSS121015 past maximum points is shown with a solid (green) line.}
\label{Mb_fig}
\end{figure*}

Taking into account the redshift, the observed V light curve shown in Sect. \ref{lc} has been translated, after K-correction \citep{kim96}, to a B light curve in the host galaxy rest frame (see Figure \ref{Mb_fig}). With the adopted distance modulus and extinction, the absolute magnitude at maximum is $M_{\rm B} = -22.6\pm0.1$ for CSS121015, which makes it one of the most luminous SNe ever observed. Moreover, the post-maximum rest-frame B decline is almost linear (see Figure \ref{Mb_fig}), with a rest-frame decay rate of $\sim$4.0  mag $(100 \mathrm{d})^{-1}$, typical of type II-Linear SNe \citep{pat94}, or a rapid decline supernova following \citet{arc12}.

Figure \ref{Mb_fig} shows a comparison of the B-band light curve of CSS121015 with  the other super-luminous, hydrogen-rich transients, namely SN 2008es \citep{gez09} and SN 2008fz \citep[][Benetti et al. in preparation]{dra10,agn10}. We also show the type IIn SN 2005gj that is spectroscopically very similar to the other SNe of the sample. We stress that SN~2005gj has been explained in terms of an interacting SN Ia \citep{pri07, ald06}. However this interpretation has been questioned by \citet{ben06,tru08,ins13b} for this class of objects, since a scenario involving the interaction between an energetic SN Ic and a dense CSM could also explain the observations. 

The shape of the $M_{\rm B}$ light curve of CSS121015 is very similar to that of SN 2008es, whose rising branch is, in this band, less constrained. However, the ROTSE unfiltered light curve indicates that the rising of SN 2008es is much steeper than that of CSS121015. Instead, the SN 2008fz light curve is slightly broader. Due to its slower evolution, the B-band luminosity of SN 2008fz matches that of CSS121015 at about 30 days after maximum. The common feature of these light curves is that the rise to maximum is steeper than the post-maximum decline, which is linear without clear inflections.
SN 2005gj is fainter by about two magnitudes than the other three SNe, and its light curve shows a steep rise and a much slower decline.

In canonical SNe IIn, the interaction with a relatively dense CSM sustains the luminosity of the supernova until late phases, making the luminosity decline rate much slower than that expected from radioactive $^{56}$Co decay \citep[e.g., SN 1988Z,][]{mt88z,kie12}. In other SNe~IIn, e.g. SNe 1994W \citep{js94w} and 2009kn \citep{ek09kn}, the late-time decline is, to a first order approximation, consistent with $^{56}$Co decay. 
The fast early decline of H-rich SLSNe (SLSNe-II) is reminiscent of SNe IIL, some of which also show signatures of early interaction, e.g. SN~1994aj \citep{sb94aj}, SN 1996L \citep{sb96l}, or of SNe IIn with fast photometric evolution, e.g. SN 1998S \citep{leo00,fas00,anu01,fas01}.  

\subsection{Physical interpretation of the explosion} 

We have presented a photometric and spectroscopic study of CSS121015, which is among the most luminous SNe ever discovered. We have highlighted photometric similarities and differences with respect to a selected sample of  SLSNe-II.\\
As mentioned in Sect. \ref{int}, different scenarios have been proposed to explain the physical outputs of SLSNe, including pair-instability explosions that produce large Ni masses \citep[e.g.][]{gal12}; the energy deposited into a supernova ejecta by a highly magnetic (B $\sim 5 \times 10^{14}$ G) neutron star spinning with an initial period of P$_i \sim  2-20$ ms \citep[a magnetar,][]{kas10}; the accretion onto a compact remnant \citep{dex13}; and the interaction of the ejecta with circumstellar material previously lost by the progenitor \citep{che11}.

First, we explore the possibility that CSS121015 and similar objects may be powered by the spin-down of a magnetar, as proposed for 
SLSNe Ic \citep[e.g.][]{ins13,nic13}. For this goal, we fit the CSS121015 pseudo-bolometric/BB light curves using magnetar models as presented by  \citet{ins13} and \citet{nic13}. The fit is initialized by comparing the observed light curve to a grid of magnetar-powered synthetic SN light curves. An exact fit is then found by $\chi^2$ minimization.

\begin{figure}
\includegraphics[width=9cm,angle=0]{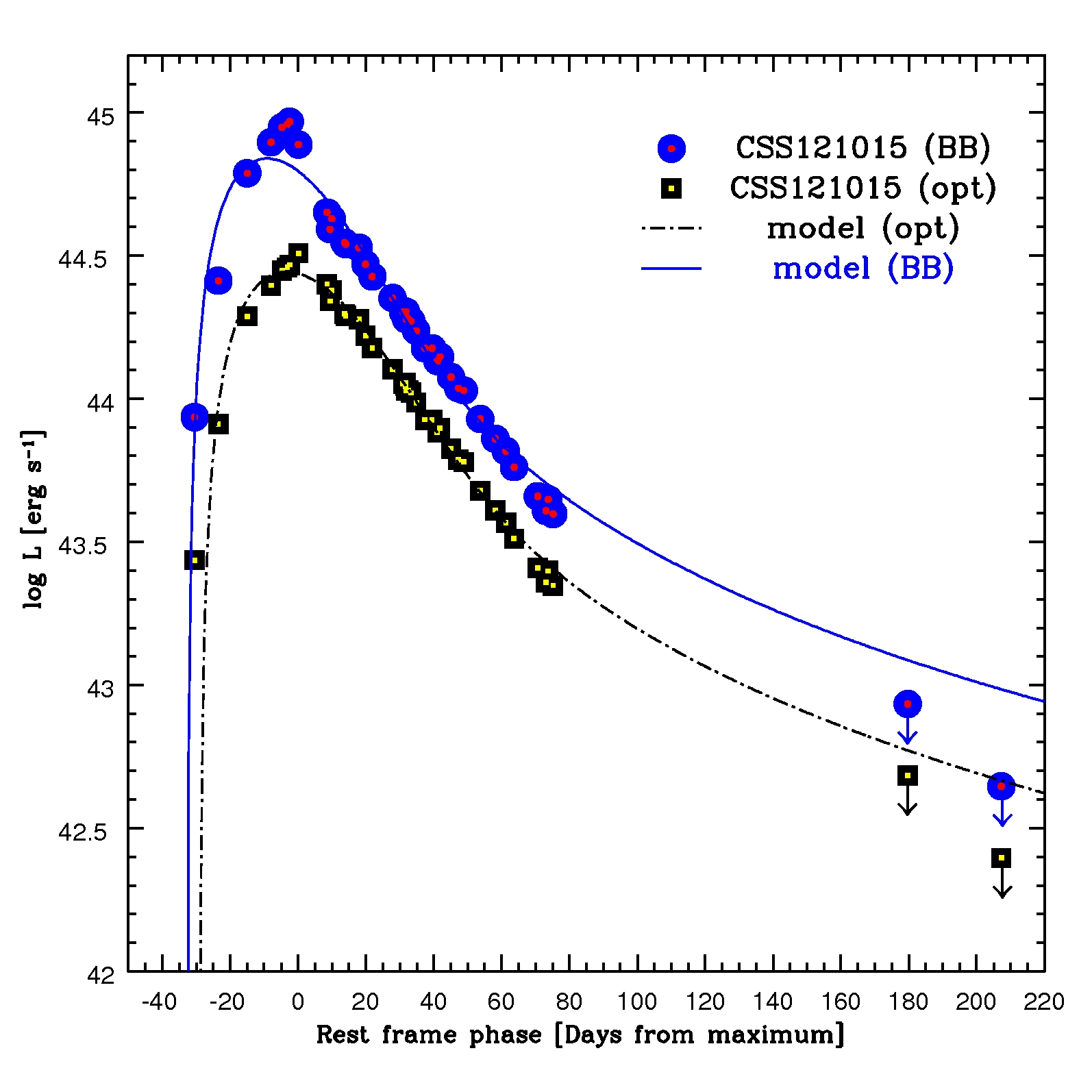}
\caption{Comparison among the pseudo-bolometric light curve of CSS121015 obtained integrating the optical contribution only, the black-body bolometric curve of CSS121015, and the two magnetar models described in the text that well reproduce the 2 curves.}
\label{bolom_model}
\end{figure}

In Figure \ref{bolom_model} we show reasonable fits to the pseudo-bolometric and BB-based 
bolometric curves of CSS121015 (Sect. \ref{bol}) with magnetar models obtained adopting the following  parameters. For the model fitting the pseudo-bolometric light curve, the input parameters were: an ejected mass of $\sim 5.5$ \M, a magnetic field of B$=2.07\times 10^{14}$ G, and a neutron star with a period of P$_i \sim  1.99$ ms.  The BB bolometric light curve was instead fitted by a model obtained with an ejected mass of $\sim 5.6$ \M, magnetic field of B$=1.42\times 10^{14}$ G, and a neutron star with a period of 
P$_i \sim  1.33$ ms \cite[see][for details on the assumptions of the model]{ins13}. While the magnetar model fits the pseudo-bolometric light curve fairly well, the model fitting the BB bolometric curve provides a worse result, since there is a poor match of the curve
before and around maximum light, where the model is somewhat broader than the extrapolated points. Remembering the uncertainties of the BB extrapolation, this fit is still quite reasonable.

Both models give a late-time light curve tail (phase \textgreater100 d) with a higher luminosity than that inferred from the observed detection limits. 
This may be a consequence of the approximations adopted in the model, (e.g. perhaps the magnetar energy is no longer fully trapped at late phases), or may be evidence that magnetar spin-down is not able to account for the full light curve of CSS121015. However, within the uncertainties of this kind of modelling, particularly in how the energy is deposited in the ejecta, a magnetar scenario cannot be excluded.

On the other hand, the narrow lines with decreasing fluxes seen in our spectra provide clear evidence of the presence of CSM lost by the progenitor before explosion. Therefore it seems logical to investigate a scenario where the luminosity is powered by interaction of the SN ejecta with this CSM, and to verify consistency with the expected evolution of massive stars. Of course, it may also be possible that the narrow lines arise from a weak interaction with low-density CSM, while some other source, such as a magnetar, powers the continuum luminosity.



We begin assuming that the progenitor of CSS121015 was a restless, very massive star ($M> 50 $\M), that during nuclear burning ejected several solar masses of material in distinct outbursts, resulting in a number of massive circumstellar shells \citep[see e.g.][]{pas10,fol11,smi11}. When the star experiences a subsequent major outburst, or the final core collapse explosion, the fast expanding ejecta soon collide with the innermost massive shell, and a fraction of the kinetic energy of the ejecta is thermalized in the shell; this is the standard picture of ejecta-CSM interaction invoked for SNe IIn.

The shock occurs at the inner boundary of the massive, optically thick CSM shell, and in most cases will be observable only indirectly. In typical SNe IIn, the CSM is optically thin and the violent collision results in a number of phenomena, particularly strong X-ray and radio emission, strong optical emission lines with multiple components, including narrow features with velocity typical of the surrounding CSM, and intermediate features with FWHM of a few thousand \kms, arising from the shocked ejecta. The scenario can account for intermediate cases such as SN 2005gj, where the CSM is thin enough to show multiple-component emission lines, but not strong X-ray and radio emission \citep{ald06}.

In the shock, kinetic energy is converted to radiation, and the shell is heated and accelerated. The luminosity rise time depends mainly on the radiation diffusion timescale in the shell (for CSS121015, we find a relatively long $\sim 40$ day rest-frame rise time), while the emerging spectrum is well fitted by a blackbody at high temperature. If the shell is initially at a large distance from the star, the early adiabatic losses after collision are small, and super-luminous peak absolute magnitudes ($M <-21$) are possible \citep{qui11}, provided the shell is sufficiently massive and opaque to efficiently thermalize a large fraction of the ejecta kinetic energy. This is the case applicable to CSS121015: at maximum light, we find that the still-opaque shell has a blackbody radius of $\sim 150$ AU (deduced from log$\,L_\mathrm{UBVRIz}\sim 44.5$ dex, assuming a BB temperature of $\sim 17300\,^{\mathrm{o}}$K, see Tab. \ref{rad}). On the other hand, the radius determined from the BB-extrapolated luminosity (see Sect. \ref{bol}) is $\sim 300$ AU which should be considered as an upper limit.

Similar opaque-shell models for SLSNe-II 2006tf and 2006gy had blackbody radii and temperatures of 300 AU and $7800\,^{\mathrm{o}}$K, and 320 AU and $\sim10000\,^{\mathrm{o}}$K, respectively \citep{smi08,smi07}. Perhaps more relevant is the comparison with SN 2008es owing to its observed similarities with CSS121015. \citet{mil09} found a temperature of $\sim15000\,^{\mathrm{o}}$K and a radius of $\sim$260 AU shortly after peak, very similar to our estimates for CSS121015.

At times longer than the diffusion timescale, radiation leaks out at a significant rate from the cooling shell between the forward and reverse shocks. As the shell cools, instabilities in the shocked region lead to clumping \citep{smi08b}. The optical depth in the rarified inter-clump regions drops rapidly, reducing the mean opacity, and finally allowing radiation from the shocked ejecta of the SN itself to escape \citep{smi08,agn10}. About seventy days after maximum, the spectra start to be dominated by broad metal features (see Fig \ref{spec_evol}), consistent with underlying Type Ic supernova shocked-ejecta. These features are relatively weak compared with that of a normal SN Ic, because in this case they appear superimposed on the extremely luminous blackbody emission from the pseudo-photosphere. These features are also very similar to those seen in SLSNe with no hydrogen in their spectra (SLSNe-Ic, cfr. Sect. \ref{Hpoor}).

\begin{table*}
\caption{Radii deduced for CSS121015} \label{rad}
\begin{tabular}{rlrlll}
\hline
\hline
phase$^*$   & log$L_{UBVRIz}^{**}$& T                           &logL$_{BB}^{**}$&R$_{UBVRIz}$&R$_{BB}$ \\
(days)         &                              &($^{\mathrm{o}}$K)&                  &      (AU)          & (AU)         \\
\hline			    		                    
-4               &44.45                          & 17700  & 44.93              &134                 &  232            \\
-2               &44.47                          & 17300  & 44.93              &144                 &  245            \\
+10            &44.45                          & 11700   & 44.71              &283                 &  391            \\
+14            &44.24                          & 10300   & 44.50              &311                 &  421            \\
+39            &43.93                          &  8400    & 44.15              &327                 &  424            \\
+47            &43.79                          &  6700    & 44.04              &438                 &  586            \\
+66            &43.44                          &  5500    & 43.82              &434                 &  672            \\
%

\hline
\end{tabular}

* - rest frame phase relative to the estimated epoch of the V maximum (JD = 2456226).\\

** -the steps followed for the computation of L$_{UBVRiz}$ and L$_{BB}$ are described in Sect. \ref{bol} 

\end{table*}

The blackbody radius increases constantly in time, as expected for an expanding and cooling pseudo-photosphere. It reaches an extension of about 450 AU by the time of our last spectrum, or about 700 AU in the case of BB extrapolation. (see Table \ref{rad}). Taking into account that the difference between the photospheric radius at the time of our last spectrum and the radius deduced at the epoch of our first spectrum is $\sim 4.5\times 10^{15}$ cm (or $\sim 6.6\times 10^{15}$ cm in the BB extrapolation), and that the time lapse between these two spectra is 70 days in the rest frame, we deduce for the pseudo-photosphere an average expansion velocity of $\sim7400$ \kms ($\sim 10600$ \kms~in the BB extrapolation). These numbers are comparable to the expansion velocity derived from the broad \Ha~emission. 
 
In this scenario, the narrow-to-intermediate width Balmer emissions that are signatures of CSM interaction could be weak (as observed in CSS121015) or entirely absent, if the CSM shell is opaque and the shock encounters no further CSM at larger radii. As stated previously, the condition of a highly opaque shell must hold, in order to generate the observed continuum luminosity \citep{smi07}. The fact that the spectrum remains featureless for a long time also demonstrates that there is a very high optical depth between the observer and the underlying supernova shocked ejecta. An interesting question is whether there are regimes of density/temperature/opacity where the intermediate-width Balmer lines from the shocked region are absent, i.e. can this scenario also describe SLSNe-Ic? While this seems plausible, detailed spectral modelling is necessary for a confirmation.

The FWHM of the broad Balmer lines are not seen to decrease significantly with time (see Sect. \ref{sp_evol}). This is probably because, due to the $\ga$ 30-40 day diffusion time in the shell, a great deal of CSM mass has been swept up by the forward shock before we observe any emission line. Deceleration of the shock likely does occur at early times, following the ejecta-CSM collision. However, by the time the transient becomes visible, the massive shell coasts at almost constant velocity, as it has been by now mostly swept up by the forward shock. The same phenomenon was observed in SN 2006tf and, according to \citet{smi07}, the fact that the shell does not slow down while it radiates $\ga 10^{51}$ erg shows that it must comprise at least several solar masses.

The narrow components of the Balmer lines arise from an unshocked, low-velocity wind external to the dense shell \citep{smi08}, which may be excited or ionised by radiation from the shell. The presence/density of this wind, perhaps along with the composition of the underlying ejecta (Type I vs Type II), is likely the origin of the spectral differences between CSS121015-like and 2006gy-like SLSNe-II, the last one displaying prominent, multi-component Balmer lines \citep{smi07}.

Since the radius deduced for CSS121015 exceeds that of typical red supergiants \citep{smi01} by two orders of magnitude, we propose that the opaque CSM shell was ejected some time prior to the SN explosion and is not bound to the star. A similar model was suggested for SN 2008es for which the CSM mass was estimated between $\sim 5$ \M \citep{mil09} and $\sim 2.5-3$ \M \citep{cha13}.
 
Following the formalisms of \citet{qui07} and \citet{smi07} for the radiation emitted by a shocked, thermalized shell, the peak luminosity is $L\propto\frac{1}{2}M_\mathrm{sh}v^2_\mathrm{ph}/t_\mathrm{max}$, where $M_\mathrm{sh}$ is the mass of the CSM shell, $v_\mathrm{ph}$ is the velocity of the pseudo-photosphere and $t_\mathrm{max}$ is the rise time of the bolometric light curve. 
If we assume that CSS121015 has a rise time $\sim 1.9$ times that of SN~2008es, a peak brighter by 0.12 dex and a comparable photospheric velocity (see Fig. \ref{spec_conf}), we estimate (with the above formula) a CSM mass of $\sim 8.5$ \M, where we have used a mass of $\sim 2.7$ \M\/ for the SN 2008es CSM shell \citep{cha13}.

We can apply a consistency check on the energetics of our model. \citet{smi07} note that, even with high efficiency in thermalizing the ejecta kinetic energy in the shell, momentum conservation tells us that the kinetic energy of the now-accelerated, shocked shell must be at least of the order of the thermal energy deposited. The thermal energy in the shell is eventually observed as radiation; integrating our pseudo-bolometric light curve implies this is $\ga 1.2 \times 10^{51}$ erg. Taking our estimated shell mass of $\sim 8.5$ \M~and velocity of 7400 \kms, we estimate a kinetic energy of $\sim 4.7 \times 10^{51}$ erg, so the parameters we have derived do seem to be consistent. This also suggests that the explosion energy of the supernova was $\ga 5-6 \times 10^{51}$ erg.

With the derived mass, we can explain the relatively fast rise and decay observed for CSS121015, compared to other luminous, slowly-evolving SNe \citep[e.g. SN 2006gy,][]{smi07,ofe07,agn09}, since the radiation diffusion time for this SN is much shorter than in those events. If we assume the mass comprising this shell was lost in a steady wind, we can estimate the mass loss rate as follows: we take the blackbody radius at maximum light ($\sim 2 \times 10^{13}$ m) as a representative radius for the CSM shell (as the shell is optically thick, the photosphere should be near the outer edge, so this is a reasonable approximation), and a wind velocity of $10-100$ \kms (where the upper limit corresponds to a typical LBV wind), and find that the wind must have began $\sim 6-65$ yr before explosion. The corresponding mass loss rate is then $\sim 0.1-1$ \M~yr$^{-1}$, which is much larger than any known stellar wind. Normal wind-driven mass loss therefore seems unlikely.


 If instead, the mass loss occurred in an outburst like those seen in SNe 1994aj \citep{sb94aj} and 1996L \citep{sb96l}, where the CSM shell had velocities of $\la 800$ \kms~(a similar velocity is also deduced from the barely resolved CSS121015 narrow lines), then the impulsive mass ejection would have happened only 1 year before the core collapse. This violent mass loss would have probably given rise to a pre-explosion optical transient, similar to those reported by \citet{pas07} for SN 2006jc, by \citet{ofe13} for SN 2010mc and by \citet{fra13b} for SN 2011ht. The inspection of the pre-outburst light curve of CSS121015\footnote{http://voeventnet.caltech.edu/feeds/ATEL/CRTS\-/1210151120044133047.atel.html} does not show any strong activity within 7 years before the explosion. However, the deepest limiting magnitudes of the pre-CSS121015 measurements are $\la 20.6$, which corresponds to an absolute magnitude of $\la -20.6$ mag. A massive ejection like those seen in SN 2006jc and proposed for SNe 1994aj and 1996L would have been well below the detection limit.

If we compare the luminosity upper limit at $\sim 178-204$ days (rest frame) with the bolometric luminosity  of SN 1987A at a similar phase, we derive an upper limit of $\la 2.5$ \M~ for the $^{56}$Ni mass ejected in the explosion. This value for the $^{56}$Ni mass is still consistent with the lower limits of $^{56}$Ni foreseen for some pair-instability scenario models \citep{wha13} in stars with initial masses between 150-200 \M.

In summary the observations of CSS121015 are consistent with a scenario in which the high luminosity arises from kinetic energy thermalized in the shock between the ejecta and a dense shell. This seems to require a circumstellar shell of a few solar masses, and therefore a massive star as its progenitor. The opaque shell implies that we do not get a direct view of the SN ejecta until approximately 60 days (rest frame) after explosion. The lines that eventually appear are reminiscent of SLSNe-Ic. For the explosion mechanism, there is no evidence to support a pair-instability SN with an high production of $^{56}$Ni mass, while a pair instability with a low $^{56}$Ni production explosion \citep{heg03,yoo12} can still be possible, but it would fail to explain the peak luminosity. On the other hand, an energetic core-collapse explosion from a stripped-envelope progenitor could be the more natural candidate for the explosion mechanism.

\subsection{H-poor SLSNe (SLSNe-Ic) in the context of CSS121015} \label{Hpoor}

We argued that the reprocessing of kinetic energy into radiation via ejecta-CSM interaction would satisfactorily explain the properties of SLSNe-II. However, whether this mechanism may also explain the behaviour of very luminous H-poor events is still a subject of debate. 
Thermonuclear explosion, triggered by pair-production in an extremely massive star, has been proposed to explain the H-poor SLSN 2007bi \citep[][but see \cite{dessart12}]{gal09}, whose late-time light curve has a slope which is consistent with $^{56}$Co decay. Nonetheless, a more canonical core-collapse, with the ejection of a relatively large amount of $^{56}$Ni, may reproduce equally well the high luminosity observed in this SN and its overall early time spectro-photometric properties  \citep{you10,mor10}. 

Another group of SLSNe-Ic shows relatively narrow light curves, with a linear decline at late-times, much faster than that expected from the $^{56}$Co decay \citep{pas10b,qui11,cho11,bar09,lel12,cho13,lun13,ins13}. This implies that these events cannot be powered by large masses of radioactive material. Indeed based on data of two recent SNe spectroscopically similar to SN~2007bi, \citet[][see also \citet{matt14}]{nic13} argued that they that are not consistent with very massive, nickel-rich ejecta.

Here we want to emphasize the similarity between CSS121015 and some SLSNe-Ic. In this view it is interesting to note that most SLSNe-II are observed in similar host dwarf galaxies as SLSNe-Ic \citep{nei11}, although SN 2006gy exploded in NGC 1260 --  a much more massive, redder galaxy.

\begin{figure}
\includegraphics[width=9cm,angle=0]{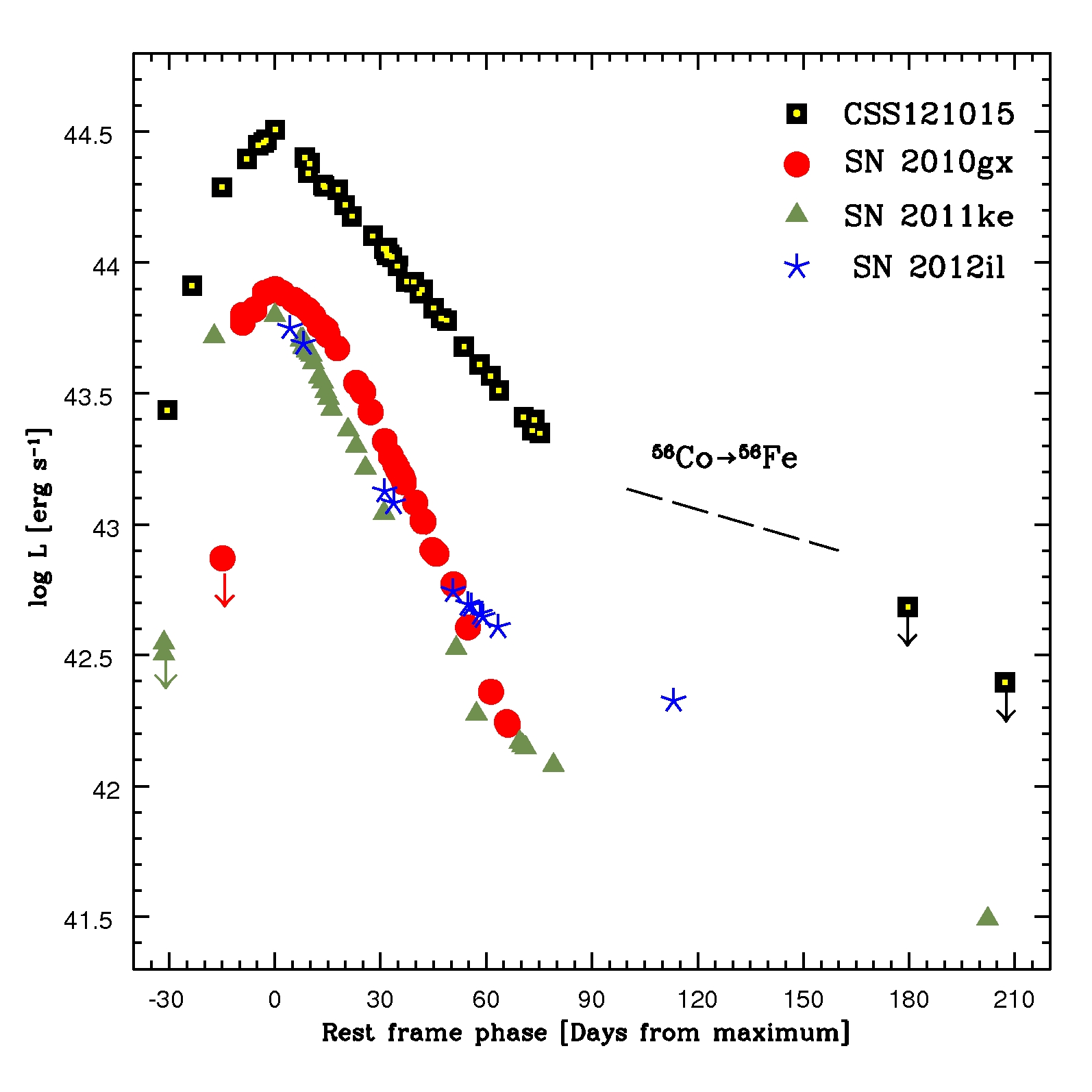}
\caption{Comparison of the quasi-bolometric light curves of CSS121015 and the SLSNe-Ic 2010gx \citep[z $\sim 0.23$, E(B-V) = 0.04 mag, from][]{pas10b};  2011ke \citep[z $\sim 0.143$, E(B-V) = 0.01 mag, from][]{ins13}; and 2012il \citep[z $\sim 0.175$, E(B-V) = 0.02 mag, from][]{ins13}. The curves have been corrected for time dilation.} 
\label{bolom_conf}
\end{figure}

To emphasize the similarity we constructed the quasi-bolometric light curves for a selected sample of H-poor SLSNe: SN 2010gx \citep{pas10b}, SN 2011ke and SN 2012il \citep{ins13} and compared them with that of CSS121015, after correction for time dilatation, in Fig. \ref{bolom_conf}. All pseudo-bolometric light curves were computed following the same prescriptions as reported in Sect. \ref{bol}.

\begin{figure}
\includegraphics[width=9cm,angle=0]{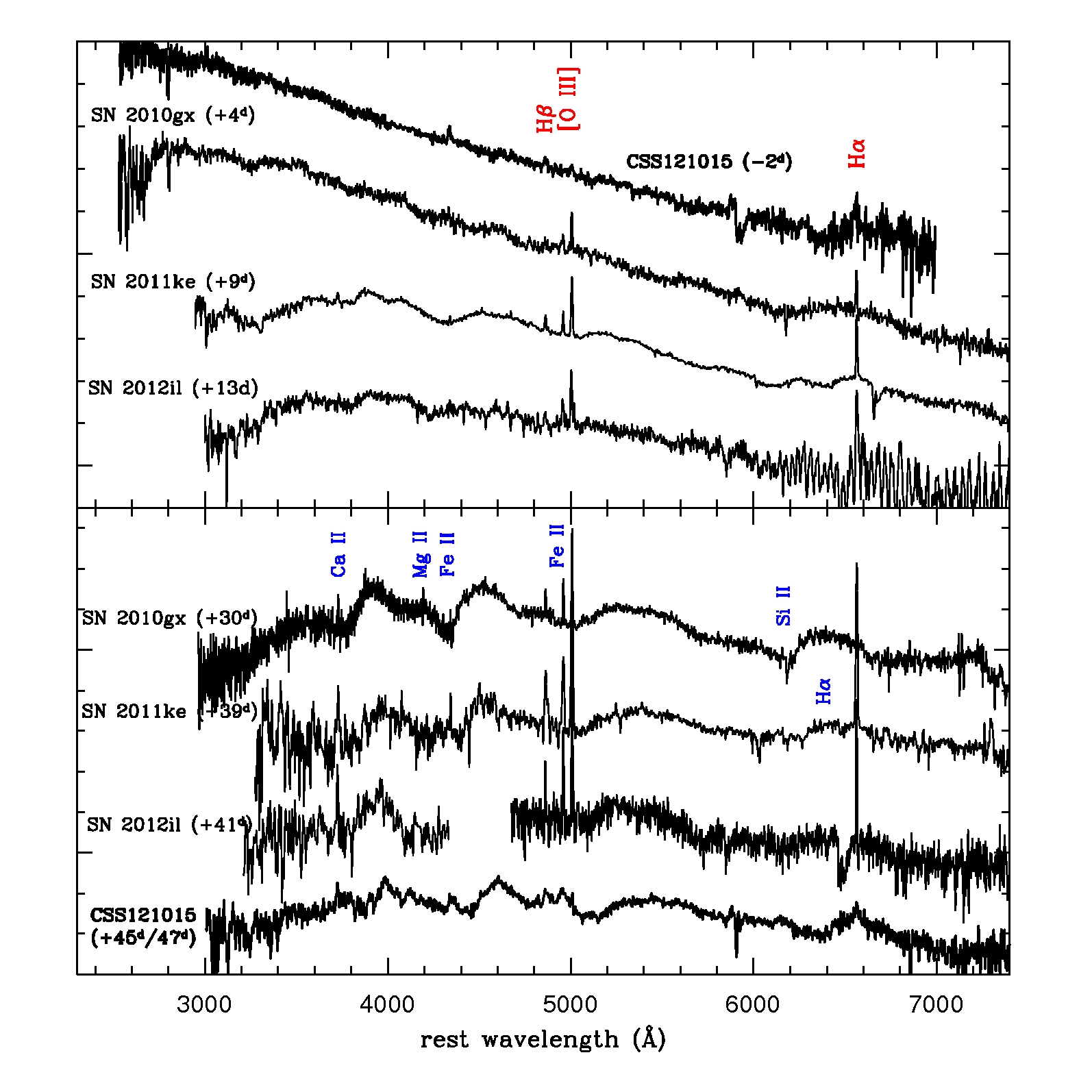}
\caption{Rest frame comparison between CSS121015 and SLSNe-Ic spectra at two phases: around the maximum light (top) and about 1-1.5 months after the maximum (bottom). 
The spectra are very similar blue-ward of 6000 \AA. The major difference is the \Ha~region, which in SLSNe-Ic is probably dominated by the Si-II 6355 \AA~transition. The expansion velocities seem to be similar in the entire sample. Line identifications are taken from \citet{pas10b}}
\label{spec_conf_pasto}
\end{figure}

In Figure \ref{spec_conf_pasto} we compare the spectra at  around the maximum light in the top panel, while spectra at phases of about 1-1.5 months after maximum are compared in the bottom panel. The phases have been computed after correcting for time dilation. A suggestive degree of similarity is seen -- particularly at 1-1.5 months, when a number of common lines are detected, including Ca II, Mg II and Fe II. The overall spectral shape is also surprisingly similar in our sample spectra, with the only difference lying in the broad feature at 6300-6600 \AA. This has been identified as either H$\alpha$ or Si II $\lambda$6355 \AA, depending on the case. In some objects (e.g. SN 2011ke), the identification of the broad feature may be disputable, with \Ha~probably being a more solid identification for this SN. However, while the broad \Ha~has strengthened in CSS121015 from phase +38 to +86 days, the corresponding feature in SN~2011ke has weakened with time.

On the other hand, the spectra around maximum light (top panel) may show significant differences, as all of the H-poor SLSNe exhibit broad lines a few days either side of maximum, while CSS121015 shows little sign of such features until at least two weeks later. The persistent featureless spectrum of CSS121015 is consistent with a highly optically thick shell, and is further evidence in favour of circumstellar interaction. But the appearance of broad (SN?) lines early in the evolution of SLSNe Ic (\citet{qui11} and \citet{nic13} have found broad lines 2-3 weeks before maximum light) may be difficult to reconcile with a model where opaque shells surround the SNe: these shells should conceal any spectral lines originating in the ejecta, until the CSM has had time to cool radiatively and decrease its optical depth. However, since many factors (e.g. CSM clumping) affect the emergent spectrum and timescales in a scenario as complex as ejecta-CSM interaction, it is difficult to say how strong a constraint this is without detailed spectral models. 

Regardless, the overall spectral similarity suggests that the underlying ejecta composition is consistent across our sample, i.e. the supernovae are probably all from stripped-envelope progenitors. If all these SNe were powered by interaction, the presence or absence of H in the spectrum could be simply a consequence of different shell properties -- viz. in SLSNe-Ic, the CSM may be H-deficient. We note that all SLSNe-Ic develop type Ic features on much longer time-scales than normal SNe Ic \citep{pas10b}, and the presence of an initially highly-opaque shell could potentially explain the cause.

\bigskip

Since CSS121015 seems to be largely consistent with the ejecta-CSM interaction scenario, we propose that this should be considered as an efficient mechanism for generating the enormous luminosities observed in many SLSNe, independent of the composition (H-rich or -deficient) of their outer stellar envelopes. However, we note that other SLSNe-Ic, including the SN 2007bi-like group, have quite different spectro-photometric properties. Therefore, it is premature to claim a common powering mechanism for {\it all SLSNe}.\\
The diversity in the spectra between different sub-types (SLSNe-Ic and SLSNe-II, CSS121015-like and 2006gy-like) may perhaps be explained in terms of differences in the structure and hydrogen content of their circumstellar cocoons.

A thorough comparison between different powering mechanisms in SLSNe II and Ic will be the topic of an in-depth, forthcoming investigation (Nicholl et al, in preparation).

\section{Conclusions}

We have presented extensive data for CSS121015, covering over 200 days (rest frame) of evolution. The collected observations, which consist of UBVRI$griz$ photometry and low-resolution optical spectroscopy, make this one of the most comprehensive data sets for a SLSN.

The analysis of the observations of CSS121015 shows that this is among the most luminous SNe ever discovered. Its photometric evolution is characterized by a relatively fast rise to maximum ($\sim40$ days in the SN rest frame), and by a post-maximum decline typical of SNe II-Linear. Its light curve shows no sign of a break to an exponential tail.

Compared to others SLSNe, the spectral evolution is relatively fast after maximum. The first available spectrum shows a very hot ($T_\mathrm{BB}\sim 17700~^{\mathrm{o}}$K) and featureless continuum. The continuum cools down quickly after maximum, and the spectra show the first broad features two weeks after maximum (in the rest frame). A broad \Ha~line is first detected at $\sim +39$ days (rest-frame). Narrow, barely-resolved Balmer and [O III] 5007~\AA~lines with decreasing flux are visible along the entire spectral evolution, and were gone in the spectra of the CSS121015 host taken about 1 year after the explosion.

The spectra are very similar to other SLSNe-II and also to SN~2005gj, previously classified as a SN IIa (Ia interacting with H-rich CSM). 

Although our analysis does not rule out magnetar spin-down as a viable mechanism to explain the enormous luminosity of SLSNe, our preferred model to explain CSS121015, and perhaps many other SLSNe (with and without H) is the interaction of the ejecta with a massive, extended, opaque shell, lost by the progenitor decades before the final explosion, which could either be an energetic core collapse or a pair instability with a low $^{56}$Ni production explosion.

\bigskip
\noindent
{\bf ACKNOWLEDGMENTS}\\
We thank the anonymous referee for very useful comments and suggestions.\\
S.B., E.C., A.P., L.T. and M.T. are partially supported by the PRIN-INAF 2011 with the project ÒTransient Universe: from ESO Large to PESSTOÓ.
Research leading to these results has received funding from the European Research Council under the European Union's Seventh Framework Programme (FP7/2007-2013)/ERC Grant agreement n$^{\rm o}$ [291222] (PI S.J.S) and  EU/FP7-ERC grant n$^{\rm o}$ [307260] (PI A.G.-Y.).  A.G.-Y. is also supported by ``The Quantum UniverseÓ I-Core program by the Israeli Committee for planning and funding and the ISF, a GIF grant, and the Kimmel award". N.E.R. and A.M.G. acknowledge financial support by the MICINN grant AYA2011-24704/ESP, and by the ESF EUROCORES Program EuroGENESIS (MINECO grants EUI2009-04170). N.E.R. acknowledges the support from the European Union Seventh Framework Programme (FP7/2007-2013) under grant agreement n. 267251. The CRTS survey is supported by the U.S.~National Science Foundation under grant AST-1313422.\\
This work is partially based on observations collected at: 1- the European Organisation for Astronomical Research in the Southern Hemisphere, Chile as part of PESSTO, (the Public ESO Spectroscopic Survey for Transient Objects Survey) ESO program 188.D-3003;
2- the Copernico 1.82m Telescope operated by INAF - Osservatorio Astronomico di Padova at Asiago;
3- the 3.6m Italian Telescopio Nazionale Galileo operated by the Fundaci\'on Galileo Galilei - INAF on the island of La Palma;
4- the 4.3m William Herschel Telescope operated by the Isaac Newton Group of Telescope;
5- the Gran Telescopio Canarias (GTC), installed in the Spanish Observatorio del Roque de los Muchachos of the Instituto de Astrof'sica de Canarias, in the island of La Palma;
6- the Liverpool Telescope, which is operated on the island of La Palma by Liverpool John Moores University in the Spanish Observatorio del Roque de los Muchachos of the Instituto de Astrofisica de Canarias with financial support from the UK Science and Technology Facilities Council;
6- the Faulkes Telescope Project, which is an educational and research arm of the Las Cumbres Observatory Global Telescope Network (LCOGTN);
7- the Large Binocular Telescope (LBT), which is an international collaboration among institutions in the United States, Italy and Germany. The LBT Corporation partners are: The University of Arizona on behalf of the Arizona university system; Istituto Nazionale di Astrofisica, Italy;  LBT Beteiligungsgesellschaft, Germany, representing the Max Planck Society, the Astrophysical Institute Potsdam, and Heidelberg University; The Ohio State University; The Research Corporation, on behalf of The University of Notre Dame, University of Minnesota and University of Virginia.\\
This research has made use of the NASA/IPAC Extragalactic Database (NED) which is operated by the Jet Propulsion Laboratory, California Institute of Technology, under contract with the National Aeronautics and Space Administration. We thank R. Kotak for taking the WHT+ISIS spectrum of Nov. 20th, 2012.

\end{document}